\documentclass[journal]{IEEEtran}

\ifCLASSINFOpdf

\else

\fi

\usepackage{graphicx}
\usepackage{amssymb}
\usepackage{float}
\usepackage{subcaption}
\usepackage{textcomp}
\usepackage{gensymb}
\usepackage{array}
\usepackage{amsmath}
\usepackage{algorithmic}
\usepackage{xcolor}

\allowdisplaybreaks
\allowdisplaybreaks

\begin{document}

\title{Online Robust MPC based Emergency Maneuvering System for Autonomous Vehicles}

\author{
\IEEEauthorblockN{Vivek Bithar \IEEEauthorrefmark{1}, Punit Tulpule \IEEEauthorrefmark{1}, Shawn Midlam-Mohler \IEEEauthorrefmark{1}} \\

\IEEEauthorblockA{\IEEEauthorrefmark{1}Mechanical and Aerospace Engineering Department \\ The Ohio State University, Columbus, OH, USA }
}

\markboth{}%
{Shell \MakeLowercase{\textit{et al.}}: Bare Demo of IEEEtran.cls for IEEE Journals}

\maketitle
\begin{center} \textbf{This work has been submitted to the IEEE for possible publication. Copyright may be transferred without notice, after which this version may no longer be accessible.}\end{center}
\begin{abstract}
Nonlinear Robust Model Predictive Control (RMPC) provides a very promising solution to the problem of automatic emergency maneuvering, which is capable of handling multiple possibly conflicting objectives of robustness and performance. Even though RMPC gives a suboptimal solution, the key challenge in real-time implementation is that it is computationally very demanding. In this paper a real-time capable robust tube MPC based framework for steering control during emergency obstacle avoidance maneuver is presented. The novelty of this framework lies in the robust integration of path planning and path following tasks of autonomous vehicles. A simulation study showcases the robust performance improvements due to the proposed strategy over a non-robust MPC in different extreme maneuvering scenarios.

\end{abstract}

\begin{IEEEkeywords}
Robust MPC, Collision Avoidance, Autonomous Vehicle, Motion Planning, Vehicle Control
\end{IEEEkeywords}

.
\IEEEpeerreviewmaketitle

\section{Introduction}

\IEEEPARstart{T}{he} obstacle avoidance system for Autonomous Vehicles (AVs) deals with planning of emergency maneuvers to avoid the impending crash with the obstacle in the path \cite{d:123} in addition to tracking the planned path and the stabilization \cite{d:124} \cite{d:125}. Early attempts at design of path planning and tracking focused on the hierarchical MPC framework that separated the task of path planning and tracking \cite{d:39} \cite{d:8} \cite{d:7}. Authors in \cite{d:37} presented a controller design, where high-level controller used a simplified point-mass vehicle model to generate a reference path based on information about the current position of the obstacle using NLMPC and fed the path to a low -level controller introduced in \cite{d:38}. In \cite{d:39}, the authors used similar hierarchical structure for the trajectory planning and tracking tasks by applying an explicit MPC control law with 4-wheel model for the low level controller for high prediction accuracy. Another variant of this hierarchical approach is presented in  \cite{d:83} with MPC based motion planning and a PID based low level controller. Due to the use of simplified point mass models for motion planning, all these approaches reported dynamic infeasibility of the path planned at operating conditions such as high longitudinal velocity and for low road friction coefficient surfaces, leading to large tracking errors because of low-level controller prioritizing vehicle stabilization over tracking. Thus, limiting the range of operating conditions and functionality in emergency scenarios, for such frameworks. 

To address this issue, the approaches that followed focused on utilizing a single MPC based framework to combine the tasks of the stabilization, trajectory generation/re-planning and tracking. Such a control system was proposed in \cite{d:40} as a shared MPC controller for ground vehicles. The controller combined stable handling envelopes from \cite{d:35} and environmental envelopes that representedd the obstacle and road edge constraints. This collision avoidance system computed obstacle free regions of the environments by joining feasible gaps in the prediction horizon as tubes and as a result transformed the non-convex optimization problem into a set of convex problems. The authors further employed a variable time step MPC execution to reduce the computational load, and to increase the look ahead time/horizon. Stability and collision avoidance objectives typically pose conflicting requirements on MPC design due to limited computational resources on board a vehicle. For stabilization, short time steps for capturing fast dynamics is recommended whereas collision avoidance requires preferably sufficient look ahead distance to react early. The controller developed in \cite{d:40} shared a controller between the driver and the motion planning system, where the system was planned to intervene whenever it detected that the intended maneuver may result in the vehicle going outside the stability envelope. 
 
The system design in \cite{d:41} extended the framework developed by \cite{d:40} and applied it to AV. The research pointed out that the controller for an AV has the advantage of having precise knowledge of the desired path instead of simply a projection of driver intent from the steering angle history as assumed in \cite{d:40}. The authors worked to improve the vehicle modeling to take advantage of the availability of the information about upcoming path and its curvature. They employed successive linearization scheme for rear tire force generation and first order hold for linearization in the longer time step region of the controller. It continued using the safe handling and obstacle avoidance envelopes introduced in \cite{d:35}. The research presented in \cite{d:42} further developed initial approach in \cite{d:40} and \cite{d:41} by introducing a point mass longitudinal controller to the system. The model linearization was improved with refined rear tire force linearization using an average rear tire slip angle. 

The approaches discussed so far did not consider the presence of uncertainties while designing the control system. The measurement errors from the Inertial Measurement Units (IMU), model parametric uncertainties such as friction coefficients, tire models and linearized dynamic model inaccuracies are present in the real environment. To be functional in such real situations, the controller is required to be robust against them. The design of collision avoidance system thus, become a motion planning problem in Uncertain and Dynamic Environment (UDE). Mobile robots subject to nonholonomic constraints in a UDE have been studied extensively in the field of the robotics, and AVs can be considered as extension of such systems. 

\cite{d:43} introduced a set-theoretic receding horizon control algorithm for robots modelled as Linear Time-Invariant (LTI) systems subject to input and state constraints and disturbances. This algorithm computes ellipsoidal approximations of exact one-step controllable sets for the LTI systems considering all the possible obstacle scenarios that the system can encounter. Then it exploited these sets online to determine specific control actions to be applied on the system to have a collision free trajectory in an oriented graph space. The resulting framework guaranteed collision free and bounded feasible trajectory regardless of any obstacle scenario occurrence. Offline computation of the sets assumed that obstacle locations on the working area are known, but the current obstacle configuration to be unpredictable at each time instant. Due to this restrictive assumption, the pre-computed sets were comprehensive in nature to provide uniform ultimate boundedness and feasible solution at each time instant. This publication showed an offline methodology that brought obstacle configuration uncertainty in addition to system state measurement errors and disturbances in the system design.  

Work presented in \cite{d:44} also explored the computation of disturbance invariant set and proposed a bi-level NLMPC optimization strategy. Disturbance-invariant are the regions in the state-space of a system that guarantee that if the system starts in the invariant set, then the system will lie in the invariant set under bounded disturbances. The strategy introduced first executed a feed-forward control optimization using approximate disturbance-invariant sets for a general nonlinear dynamic system at an instant and next performed an optimization to propagate the ellipsoid for the updated controller and nominal trajectory for the next instant of time. It showed the framework was able to compute and execute a trajectory for a quadrotor system with a nonlinear backstepping controller in a space where the obstacles were stationary. This paper, in addition to attempting online robust MPC algorithm also attempted to provide a solution for a general nonlinear system. The work did not provide any computational performance information of the controller system for the experiment performed.

In literature, the uncertainty in the MPC framework has also been treated with Tube Based MPC Scheme (TMPC), for example, \cite{d:45}  used this structure for a general Lipschitz nonlinear systems to show that invariant sets for the system can be computed offline, and does not require recomputing the sets for every nominal trajectory provided The invariant set can be specified as an invariant space around the nominal trajectory. \cite{d:45} used offline computed control invariant sets and demonstrated obstacle avoidance by combining tasks of planning and lane keeping on straight icy roads with stability bounds designed to limit vehicle states to linear regime of tire force generation. It developed a non-linear force input Lipschitz bicycle model with the assumption of additive uncertainty and simplified rear and front tire force linear model. For robust obstacle avoidance, it projected the computed control invariant set on the obstacle position ellipsoid constraint and solved nominal MPC with updated obstacle constraints. Although the controller execution rate was slow (approx. 100 [ms]) compared to the dynamics involved due to NLMPC optimization, this did present a functional approach of applying robust MPC for developing obstacle and path tracking system of an AV. 

 Existing researches have not presented a real-time implementable control system that can arbitrate between objectives of obstacle avoidance, stabilization and path tracking, and be robust against disturbances for emergency maneuvering. In this paper we propose a robust obstacle avoidance control which integrates the tasks of path planning and path tracking. The proposed approach improves the controller performance in terms of tracking error and vehicle stability in presence of uncertainties. While ensuring the robustness, this approach addresses the controls’ path infeasibility problem. TMPC is selected as an obstacle avoidance and control system design methodology due to its efficient multiple-objective and explicit robust constraint handling capabilities. In the absence of obstacles in the perception region of the vehicle, the admissible region is restricted to a lane width around the prescribed reference path and the problem transforms to a reference path tracking. The objective of path tracking is conflicting with the obstacle avoidance objective and the implemented novel robust MPC (RMPC) system described in this study provides an effective robust solution for both problem with a single system design. The approach is employed in a simulation study to demonstrate improvements over a deterministic MPC and real-time computation capabilities. We used CarSIM and Simulink\textregistered{} co-simulation environment to test the proposed approach on three different extreme maneuvering scenarios. 



The structure of the paper is as follows. Section II introduces the mathematical models used in the proposed control system. Section III presents the TMPC based robust controller and details the design process. Section IV elaborates the simulation setup, estimation of the system disturbance, evaluates and compares the performance of the designed robust  control system with a deterministic non-robust control system in different test cases. Section V provides a conclusion and a discussion on the potential future improvements of the designed control system.

\section{Vehicle Dynamics and Tire Model} 
\noindent This section  presents the vehicle dynamics model and tire model of the control system, and presents the process of linearization and discretization for developing the Linear Time Varying (LTV) system model of the vehicle. It presents a nonlinear single track model, describing the lateral dynamics \cite{d:42}, and introduces the concept of Center of Percussion (CP) for reducing the modeling error due to linearization. Further, it introduces the tire model and  details the process of linearization and discretization of the nonlinear model using the tire model presented.

\subsection{Single Track Vehicle Model}
\begin{figure*}
    \centering
    \includegraphics[width=\textwidth, height=3in]{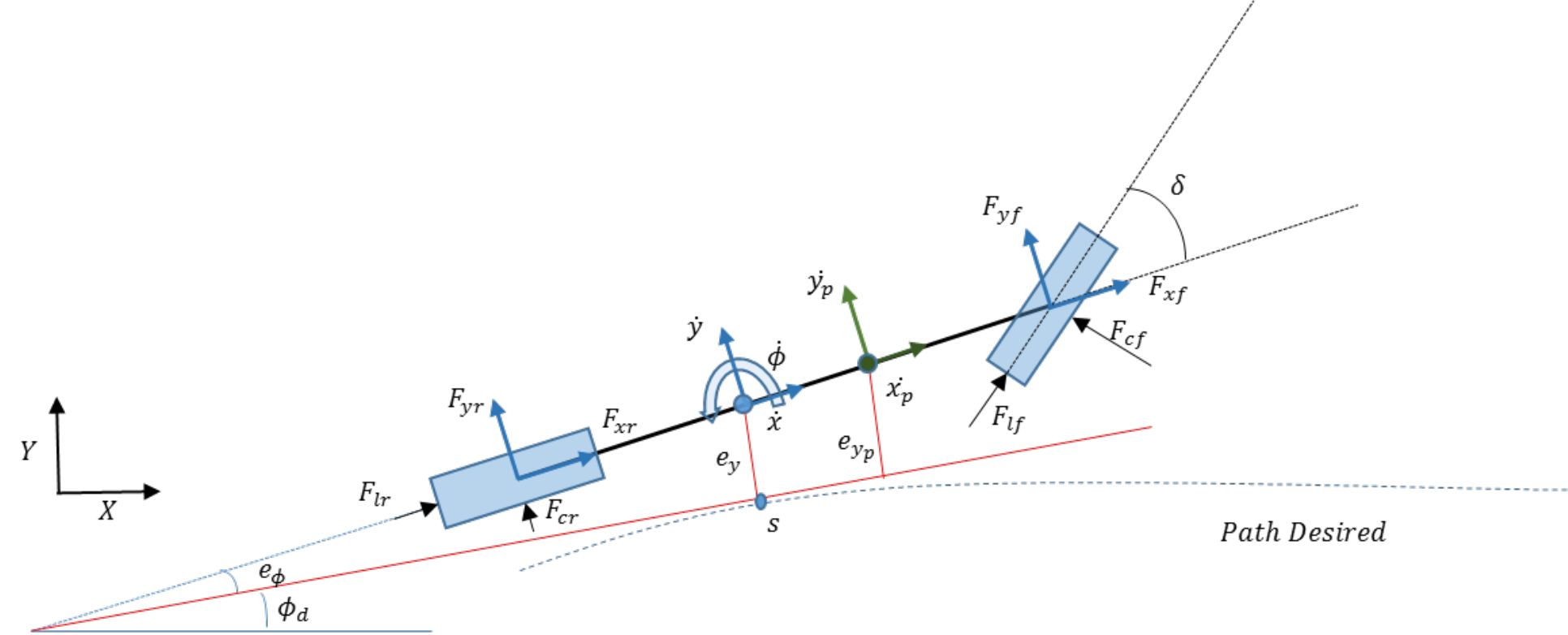}
    \caption{Single Track Vehicle Model. ($F_{xf},F_{yf},F_{xr}$ and $F_{yr}$) depict the forces in the vehicle body-fixed frame, ($F_{lf},F_{cf},F_{lr}$ and $F_{cr}$) depict forces in the tire-fixed frame, $\dot{\phi}$ depict the rotational velocity about the CG, and ($\dot{y}$ and $\dot{x}$) represent translational velocities. The desired path reference frame coordinates $e_y$ and $e_\phi$ are illustrated along with $\phi_d$. $p$ denotes the distance from CG to CP and ($\dot{y}_p$, $\dot{x}_p$) are the translational velocities about the CP.}
    \label{V_Model}
\end{figure*}

\noindent Fig. (\ref{V_Model}) shows a schematic diagram of a single track model. The force balance and path tracking error dynamics leads to the following set of differential equations to describe the vehicle motion. 

\begin{align}
    {m}\ddot{{y}} {\ } &{=-m\ }\dot{{x}}\dot{{\phi }}{+2}{{F}}_{{yf}}{+2}{{F}}_{{yr}} \label{latacc}\\ 
    {{I}}_{{z}}\ddot{{\phi }}{\ }&{= \ 2a}{{F}}_{{yf}}{-}{2b}{{F}}_{{yr}} \label{rotacc} \\ 
    {\dot{{e}}}_{{\phi }}{\ }&{= \ }\dot{{\phi }}{-}{\dot{{\phi }}}_{{d}}{\ } \label{eryawrate}\\ 
    {\dot{{e}}}_{{y}}{\ }&{= \ }\dot{{y\ }}{{cos} \left({{e}}_{{\phi }}\right)\ }{+}\dot{{x\ }}{sin}{}{(}{{e}}_{{\phi }}{)} \label{errlat}\\ 
    \dot{{s_d}}{\ }&{=}\frac{\dot{{x\ }}{cos} 
    \left({{e}}_{{\phi }}\right)\ {-}\dot{{y}}{\ sin}{}{(}{{e}}_{{\phi }}{)}}{(1-K(s_d)e_y)} \label{errdist}
\end{align}

\noindent where ${m}$ and ${{I}}_{{z}}$ denote the vehicle mass and yaw inertia, respectively, ${a}$ and ${b}$ denote the distances from the vehicle center of gravity to the front and rear axles, respectively. $\dot{{x}}$ and $\dot{{y}}$ denote the vehicle longitudinal and lateral velocities, respectively, and $\dot{{\phi }}$ is the yaw rate around a vertical axis at the vehicle's center of gravity. ${{e}}_{{\phi }}$ and ${{e}}_{{y}}$ in fig. (\ref{V_Model}) denote the vehicle orientation and lateral positions, respectively, in a path aligned coordinate frame and ${{\phi }}_{{d}}$ is the angle of the tangent to the path in a fixed coordinate frame. ${s_d}$ is the vehicle longitudinal position along  the desired path. ${{F}}_{{yf}}$ and ${{F}}_{{yr}}$ are  the front and rear tire forces acting along the vehicle lateral axis, ${{F}}_{{xf}}$ and ${{F}}_{{xr}}$ forces acting along the vehicle longitudinal axis, and $K(.)$ denotes the radius of curvature at path distance $s_d$. Equations (\ref{eryawrate}), (\ref{errlat}) and (\ref{errdist}) are from the path centered reference frame which directly provides tracking error. 
Here, the assumption is that, the steering system in the vehicle can only control the angle of front wheels.
\noindent The longitudinal and lateral tire force components are modelled as 

\begin{align}
  {{F}}_{{yf}\ }&{=}{\ {F}}_{{lf\ }}{{sin} \left({{\delta}}_{{f}}\right)\ }{+}{{F}}_{{cf}}{\ cos}{}{(}{{\delta}}_{{f}}{)}  \label{latfrontforce} \\ 
  {{F}}_{{yr}\ }&{=}{\ {F}}_{{cr}}  \label{latrearforce}
\end{align}

\noindent The steering angle is assumed small and therefore, ${{F}}_{{yf}}{\approx }{{F}}_{{cf}}$ in (\ref{latfrontforce}). Tire forces ${{F}}_{{lf}}$ and ${{F}}_{{yf}}$ are defined by the brush tire model \cite{d:76} and modified in \cite{d:77} as a function of slip angles,${\ }{{\alpha }}_{{*}}$, where $(*)$ denotes either the front ($f$) or the rear ($r$) tire model specific parameters.  

\begin{equation} 
    \begin{split}
        F_{y[f,r]} &= 
                \begin{cases}
                   \left[ \begin{array}{l}
                        -C_{\alpha_*}\tan {\alpha_* }\\ 
                        +\frac{C_{\alpha_*}}{\theta}\left| \tan \alpha_*\right|\tan\alpha_* \\
                        -\frac{C_{\alpha_*}}{3\theta^2} \tan^3 \alpha_* 
                    \end{array} \right], & \left| \alpha_* \right| < \tan^{-1}
                    {\theta}, \\ \ \ \ \ \ \ \ \ \ \, &   \theta = \frac{3\mu F_{z_*}}{C_{\alpha_*}} \\ 
                    -\mu F_{z_*}sgn(\alpha_*), &  \text{otherwise} 
                \end{cases} \\
                    &= f_{tire}(\alpha_{[f,r]}, F_{z[f,r]})
    \end{split}
    \label{brushtiremodel}
\end{equation} 

\noindent where ${\mu}$ is the tire-road surface coefficient of friction, ${{F}}_{z_*}$ is the normal force and ${{C}}_{{\alpha }_*}$ is the tire cornering stiffness. The parameter ${{C}}_{\alpha _*}$ can be estimated from the experimental ramp steer data \cite{d:58}, the coefficient of friction can be estimated online using approaches such as mentioned in \cite{d:78} \cite{d:79}. Due to assumption of constant speed in this problem setup, ${{\ F}}_{{z[f,r]}}$ has been calculated from the geometrical static weight distribution.  

The tire slip angles ${{\alpha }}_{{[f,r]}}$ are described in terms of the vehicle states and steering angle input by using small angle approximation as 

\noindent

\begin{align}
  \begin{split}
   &{{\alpha }}_{{f}\ }{=}{{\ {tan}}^{{-}{1}} 
     \left(\frac{\dot{{y}}{+a\ }\dot{{\phi }}}{\dot{{x}}}\right){-}{\delta}{\ \approx }\ }\frac{\dot{{y}}{+a\ }\dot{{\phi }}}{\dot{{x}}}{-}{\delta} \\  {\ \ } &{{\alpha }}_{{r}\ }{=}{{\ {tan}}^{{-}{1}} \left(\frac{\dot{{y}}{-}{b\ }\dot{{\phi }}}{\dot{{x}}}\right)\ }{\ }{\approx \ }\frac{\dot{{y}}{-}{b\ }\dot{{\phi }}}{\dot{{x}}}{\ } 
  \end{split}
  \label{slipangles}
\end{align}

 Clearly, from (\ref{latfrontforce}), (\ref{brushtiremodel}) and (\ref{slipangles}), it can be observed that the relationship between the front lateral force and the front slip angle is non-convex. To linearize the input-output system, the control action could be directly front lateral force command (instead of steering angle). The relationship between steering angle, which is the real input, to front lateral force depends on vehicle states, but can be extracted outside the optimization solver. A low level look-up table then can be designed to extract the steering angle from of the lateral force command and current states, or a tire model similar to (\ref{brushtiremodel}) can be utilized. In this paper, the tire model approach has been followed. 

For the rear tire , this technique cannot be applied due to lack of actuation at the rear axle (\ref{slipangles}). The rear lateral force is a function of the vehicle states and the function is a non-convex. As a result, typically a linearization of rear tire model is employed using a constant linear gain \cite{d:70}. 

\noindent 

\begin{align}
    \begin{split}
        {{F}}_{{yr}}{\ \approx }{\ {F}}_{{cr}\ }{=}{{\ C}}_{{\alpha }{r}}{{\alpha }}_{{r}\ }{=}{\ }\frac{{{C}}_{{\alpha }{r}}}{\dot{{x}}}{(}\dot{{y}}{-}{b}\dot{{\phi }}{)}
    \end{split} 
    \label{approxlatreartireforce}
\end{align}
where gain $C_{\alpha_r}$ is called cornering stiffness. 

\begin{figure}
\centering
\begin{subfigure}{0.45\textwidth}
   \includegraphics[width=0.92\textwidth, height=2in]{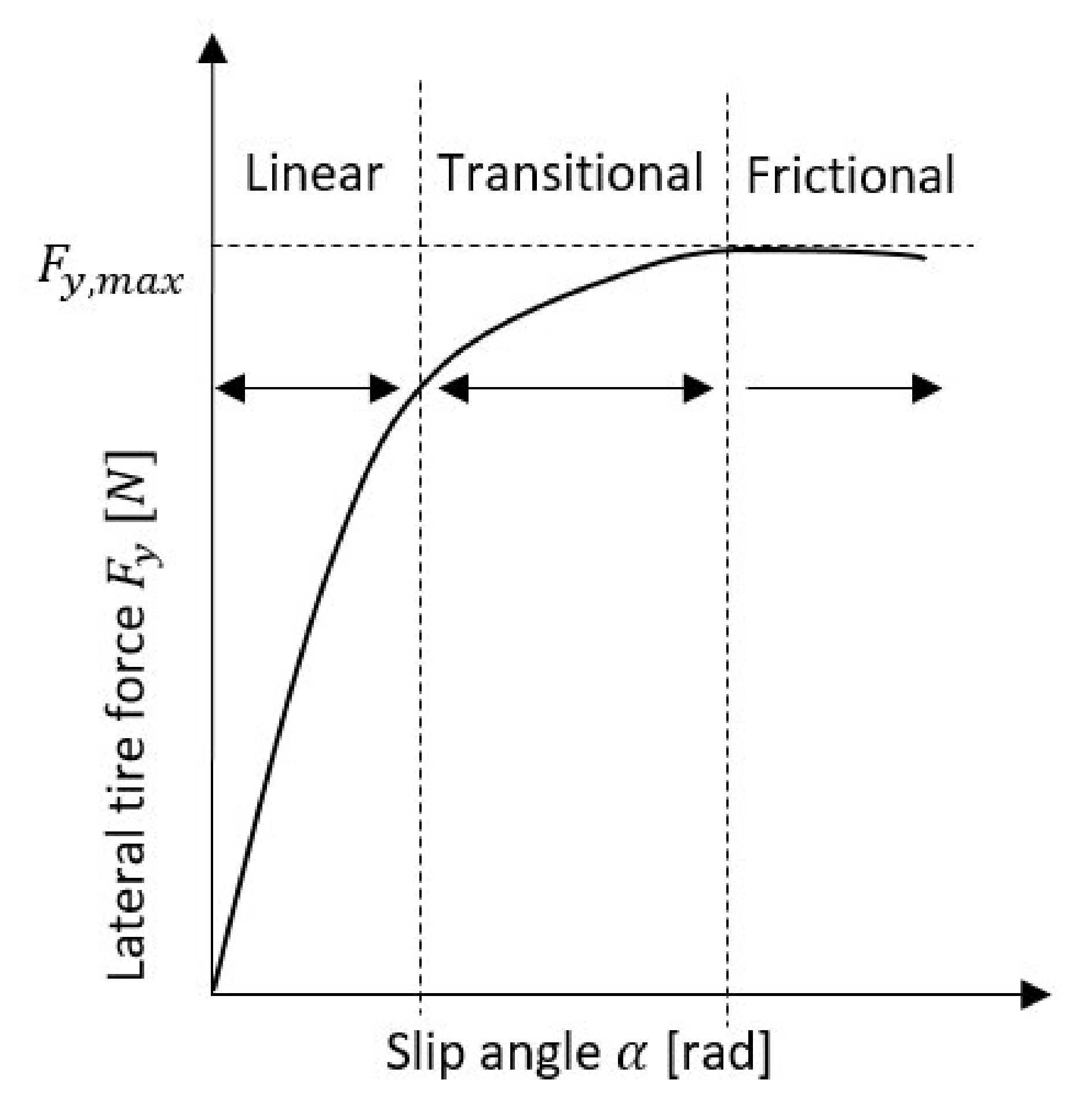}
    \caption{Characteristic curve between tire slip angle and lateral tire force.}
\end{subfigure} 

\begin{subfigure}{0.45\textwidth}
\centering
   \includegraphics[width=.90\textwidth, height=2.1
   in]{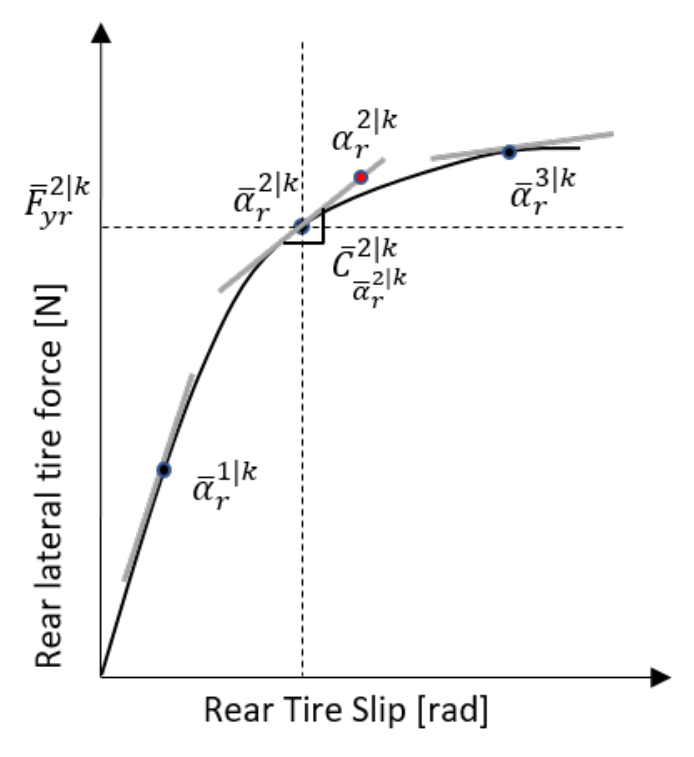}
    \caption{ Rear tire model, based on previous control step {k-1}, rear slip angles ($\bar{\alpha}_{r}^{i|k}$, i=1,2,3) predictions, represented by heavy dot marker with potential linearizations (represented by solid grey lines) at each step. The ${\alpha}_{r}^{2|k}$ depicts the current step {k}, rear slip angle solution at prediction step 2, represented by red dot marker.} 
\end{subfigure}
    \caption{Lateral tire force model}\label{figlattireforcemodel}
\end{figure}

Using a linear gain for representation of linear tire force limits the tire fidelity to the linear region of the tire force curve as illustrated by the fig. (2a). The linear gain model for the non-linear curve region can overestimate the tire force which can result in the instability of the vehicle \cite{d:67}. To reduce the tire modeling error the rear tire force can be determined by an affine linearization approach. The affine linearization is expressed in (\ref{affinelatfrontforce}) after linearization of the (\ref{brushtiremodel}) at a slip angle of ${{\ }\overline{{\alpha }}}_{{r}}$ which is estimated from the measurement of the states using (\ref{slipangles}),  where ${\overline{{C}}}_{{\overline{{\alpha }}}_{{r}}}$ is the resulting cornering stiffness at the same slip angle. In this paper $(\overline{*})$ notation represents an estimated parameter for the vehicle dynamics model based on available state measurements.

Model predictive controller computes a sequence of inputs at different prediction steps in its horizon resulting in an optimized trajectory at each controller execution step and this further facilitates determination of the rear tire force by successive linearization of the tire model (\ref{brushtiremodel}) expressed in (\ref{affinelatfrontforcestep}) where ${{\overline{{C}}}^{{i|k}}}_{{\overline{{\alpha }}}^{{i|k}}_{{r}}}$ is the local cornering stiffness estimated at a predicted slip angle ${\overline{{\alpha }}}^{{i|k}}_{{r}}$ for a prediction step${\ i}$ at the control execution step ${\ k}$. Fig. (2b) illustrates the process of determination of rear tire force model for each prediction step using the sequence of predicted slip angles ${\overline{{\alpha }}}^{{i|k}}_{{r}}$. This successive linearization of the tire model for the MPC converges within a few initial controller execution steps \cite{d:42}\cite{d:91} and although, it changes the dynamic model from time invariant to time varying, it reduces the modelling errors for lateral force estimation by accurately capturing the non-linear relationship of vehicle states and the rear tire force. 

\begin{align}
    {{F}}_{{yr}}{\ \approx }{\ {F}}_{{cr}}{\ }&{=}{\ }{\overline{{F}}}_{{yr}}{-}{\overline{{C}}}_{{\overline{{\alpha }}}_{{r}}}{(}{{\alpha }}_{{r}}{-}{\overline{{\alpha }}}_{{r}}{)} \label{affinelatfrontforce}\\
    {{F}}^{{i|k}}_{{yr}}{\ \approx }{\ {F}}_{{cr}}{\ }&{=}{\ }{{\overline{{F}}}^{{i|k}}}_{{yr}}{-}{{\overline{{C}}}^{{i|k}}}_{{\overline{{\alpha }}}^{{i|k}}_{{r}}}{(}{{\alpha }}_{{r}}{-}{\overline{{\alpha }}}^{{i|k}}_{{r}}{)} \label{affinelatfrontforcestep}
\end{align}

 Further, loss of accuracy loss due to linearization of rear tire force can be restricted by writing the equation of lateral motion (\ref{latacc}) in vehicle body reference frame with respect to Center of Percussion (CP) instead of Center of Gravity (CG) \cite{d:93}. CP, as shown in fig. (\ref{V_Model}), is a point along the vehicle longitudinal axis. As observed in (\ref{latacc}) and (\ref{rotacc}), the rear tire lateral forces have two effects on dynamics, a constant lateral acceleration on the vehicle body and an angular acceleration around the CG. At CP these two effects cancel each other out and thus, lateral velocity at CP is not influenced by the rear tire forces \cite{d:92}.  The dynamic equations with single rear tire force linearization expressed at CP are:
 
\begin{align}
        &{\dot{{y}}}_{{p}}{\ } {= \ }\dot{{y}}{+p}\dot{{\phi }}{,\ \ }{\dot{{x}}}_{{p}}{=}\dot{{x}}{ ,\ \ p=}\frac{{{I}}_{{z}}}{{mb}} \label{perclatlongvel}\\ 
        &\ddot{{{y}}_{{p}}} {\ =-\ }\dot{{{x}}_{{p}}}\dot{{\phi }}{+}
        \left(\frac{{2}}{{m}}{+}\frac{{2a}}{{mb}}\right){{F}}_{{yf}} \label{perclatvel}\\ 
    \begin{split}
        &\ddot{{\phi }}{\ }{= \ }\frac{{2a}{{F}}_{{yf}}}{{{I}}_{{z}}}{-}{2b}{\overline{{F}}}_{{yr}}{+2b}{\overline{{C}}}_{{\overline{{\alpha }}}_{{r}}}
        \left({{\alpha }}_{{r}}{-}{\overline{{\alpha }}}_{{r}}\right)\\ \ 
        &{\ \ \ = \ }\frac{{2a}{{F}}_{{yf}}}{{{I}}_{{z}}}{-}\frac{{2b}{\overline{{F}}}_{{yr}}}{{{I}}_{{z}}{\dot{{x}}}_{{p}}}{+}\frac{{2b}{\overline{{C}}}_{{\overline{{\alpha }}}_{{r}}}}{{{{I}}_{{z}}\dot{{x}}}_{{p}}}{  }
        \left({\dot{{y}}}_{{p}}{-}
            \left({p+b}\right)\dot{{\phi }}\right) \\ 
        & \ \ \ \ \ \ \ {-}\frac{{2b}{\overline{{C}}}_{{\overline{{\alpha }}}_{{r}}}}{{{I}}_{{z}}}{\ }{\overline{{\alpha }}}_{{r}}{\ } 
    \end{split} \label{percrotacc}\\
        &{\dot{{e}}}_{{\phi }}{\ }{= \ }\dot{{\phi }}{-}{\dot{{\phi }}}_{{d}}{\ } {\approx} \ \dot{{\phi }}{-}{\dot{{x_{p} }}}{\kappa(\overline{s}_d)}{\ } \label{errheadingpath}\\ 
        &{\dot{{e}}}_{{y}}{\ }{= \ }
        \left({\dot{{y}}}_{{p}}{-}{p}\dot{{\phi }}\right){\ }{{cos} 
        \left({{e}}_{{\phi }}\right)\ }{+}{\dot{{x}}}_{1{p}}{{sin} 
        \left({{e}}_{{\phi }}\right)\ }{\ }{\approx }{\ \dot{{y}}}_{{p}}{-}{p}\dot{{\phi }}{+}{\dot{{x}}}_{{p}}{{e}}_{{\phi }} \label{errlatpath}\\ 
    \begin{split}
        &\dot{{s_d}}{\ }{= \ }\frac{{\dot{{x}}}_{{p}}{\ }{{cos}
         \left({{e}}_{{\phi }}\right)}{\ }{-}
        \left({\dot{{y}}}_{{p}}{-}{p}\dot{{\phi }}\right){{sin}
        \left({{e}}_{{\phi }}\right)\ }}{(1-\kappa(s_d)e_y)} \\
        &{\ \ \ \approx } \ \frac{{\dot{{x}}}_{{p}}{-}{\dot{{y}}}_{{p}}{{e}}_{{\phi }}{+p}\dot{{\phi }}{{e}}_{{\phi }}}{(1-\kappa(s_d)e_y)}
        {\  \approx \ }\frac{{\dot{{x}}}_{{p}}{-}{\dot{{y}}}_{{p}}{\overline{{e}}}_{{\phi }}{+p}\dot{{\phi }}{\overline{{e}}}_{{\phi }}}{(1-\kappa(\overline{s}_d)\overline{e}_y)} 
    \end{split} \label{errdistperc}
\end{align}

\noindent where  ${\dot{{x}}}_{{p}}$ represents longitudinal velocity at CP and assumed to be constant and $[{\overline{{e}}}_{{\phi }},{\overline{{e}}}_{{y }},{\overline{{s}}}_{{d }}]$ is the heading error, lateral error and path distance measurements respectively for the controller execution step. $\kappa(s_d)$ represents the radius of curvature of the reference path at $s_d$ and $\overline{s_d}$ denotes the distance measurement for the controller execution step.
 
\noindent \textbf{Assumption 1:} Approximations for $\dot{e}_y$ and $\dot{{s_d}}$ assume small $e_{\phi }$. Equation (\ref{errdistperc}) assumes constant $[{\overline{{e}}}_{{\phi }},{\overline{{e}}}_{{y }},{\overline{{s}}}_{{d }}]$, hence  ${{e}}_{{\phi}}$, ${{e}}_{{y}}$, and ${{s}}_{{d}}$ preserve the linearity. 

\noindent \textbf{Assumption 2:} The signal ${\dot{{\phi }}}_{{d}}$ is known at every time step over a finite time horizon and is approximated as shown in (\ref{errheadingpath}). 

\noindent The equations (\ref{perclatlongvel}-\ref{errdistperc}) can be organized in a state space form as:
\begin{align}
    \dot{{X}}{=AX+Bu+L+w} \label{genstatespace}
\end{align}

\noindent where, ${X=[}\dot{{{y}}}_{{p}},\dot{{\phi }}{,\ }{{e}}_{{\phi }}{,\ }{{e}}_{{y}}{,s_d]}$ represents the state vector, ${w}$ represents the additive disturbance in the model, and the matrices $A$, $B$ and $L$ are shown in the following equation. 

\begin{align}
 \begin{split}
  &{A=}
  \left[ \begin{array}{ccccc}
    0 & {-}{\dot{{x}}}_{{p}} & {0\ } & 0 & 0 \\ 
    \frac{{2b}}{{{I}}_{{z}}{\dot{{x}}}_{{p}}} & {-}\frac{{2b}
    \left({b+p}\right)}{{{I}}_{{z}}{\dot{{x}}}_{{p}}} & {0\ } & 0 & 0 \\ 
    0 & {1} & 0 & 0 & 0 \\ 
    {1} & {-}{p} & {\dot{{x}}}_{{p}} & 0 & 0 \\ 
    {-}\frac{{\overline{{e}}}_{{\phi }}}{(1-\kappa(\overline{s}_d)\overline{e}_y)} & \frac{{p}{\overline{{e}}}_{{\phi }}}{(1-\kappa(\overline{s}_d)\overline{e}_y)} & 0 & 0 & 0 \end{array}
    \right] \\ \\
    &{B=}
        \left[ \begin{array}{c}
        \frac{{2}}{{m}}{+}\frac{{2a}}{{mb}} \\ 
        \frac{{2a}}{{{I}}_{{z}}} \\ 
        0 \\ 
        0 \\ 
        0 \end{array}
        \right] 
        {L=}
            \left[ \begin{array}{c}
            0 \\ 
            {-}\frac{{2b}{\overline{{F}}}_{{yr}}}{{{I}}_{{z}}{\dot{{x}}}_{{p}}}{-}\frac{{2b}{\overline{{C}}}_{{\overline{{\alpha }}}_{{r}}}}{{{I}}_{{z}}}{\ }{\overline{{\alpha }}}_{{r}} \\ 
            {-}{\dot{{\phi }}}_{{d}} \\ 
            0 \\ 
            \frac{{\dot{{x}}}_{{p}}}{(1-\kappa(\overline{s}_d)\overline{e}_y)} \end{array}
            \right]{\ }
 \end{split} \label{ABL}
\end{align}

The pair (A,B) is not stabilizable because path distance state, ${s_d}$, is not controllable by the steering input. The state is therefore treated as an augmented state in the system model and the system matrices are partitioned according to stabilizable and augmented states as shown below

\begin{align}
    \dot{{X}\ }{=}
        \left[ 
            \begin{array}{cc}
        {{A}}_{{stab}} & 0 \\ 
        {{A}}_{{aug}} & 0 
            \end{array}
        \right]{\ X+}
        \left[
            \begin{array}{c}
        {{B}}_{{aug}} \\ 
        0   \end{array}
        \right]{u+}
        \left[ 
            \begin{array}{c}
        {{L}}_{{stab}} \\ 
        {{L}}_{{aug}}
            \end{array}
        \right]{+w}  \label{bifurstatespace}
\end{align}

\noindent  
\noindent After incorporating the sequence of rear tire force linearizations (\ref{affinelatfrontforcestep}) and the sequence of state information from the previous time instant $k$, the system model can then be expressed as sequence of models:

\begin{align}
    \dot{{X}\ }{=}
        \left[ 
            \begin{array}{cc}
            {{A}}^{{i|k}}_{{stab}} & 0 \\ 
            {{A}}^{{i|k}}_{{aug}} & 0 \end{array}
        \right]{\ X+}
        \left[ 
            \begin{array}{c}
            {{B}}^{{i|k}}_{{aug}} \\ 
            0 \end{array}
        \right]{u+}
        \left[ 
            \begin{array}{c}
            {{L}}^{{i|k}}_{{stab}} \\ 
            {{L}}^{{i|k}}_{{aug}} \end{array}
            \right]{+w}  \label{bifurstatespacestep}
\end{align}

\noindent where, 

\begin{align}
    \begin{split}
        &{A^{i|k}\ =}
            \left[ 
                \begin{array}{ccccc}
                0 & {-}{\dot{{x}}}_{{p}} & {0\ } & 0 & 0 \\ 
                \frac{{2b}}{{{I}}_{{z}}{\dot{{x}}}_{{p}}} & {-}\frac{{2b}
                \left({b+p}\right)}{{{I}}_{{z}}{\dot{{x}}}_{{p}}} & {0\ } & 0 & 0 \\ 
                0 & {1} & 0 & 0 & 0 \\ 
                {1} & {-}{p} & {\dot{{x}}}_{{p}} & 0 & 0 \\ 
                {-}\frac{{\overline{{e}}}^{i|k}_{{\phi }}}{(1-\kappa({\overline{s}}^{i|k}_{d}){\overline{e}}^{i|k}_{y})} & \frac{{p}{\overline{{e}}}^{i|k}_{{\phi }}}{(1-\kappa(\overline{s}_d)\overline{e}^{i|k}_{y})} & 0 & 0 & 0 \end{array}
            \right] \\ \\
            &{{L}}^{{i|k}\ }{=}
                \left[ 
                \begin{array}{c}
                 0 \\ 
                {-}\frac{{2b}{{\overline{{F}}}^{{i|k}}}_{{yr}}}{{{I}}_{{z}}{\dot{{x}}}_{{p}}}{-}\frac{{2b}{{\overline{{C}}}^{{i|k}}}_{{{\overline{{\alpha }}}^{{i|k}}}_{{r}}}}{{{I}}_{{z}}}{\ }{{\overline{{\alpha }}}^{{i|k}}}_{{r}} \\ 
                {-}{\dot{{x_{p} }}}{\kappa({\overline{s}}^{i|k}_{d})}\\ 
                0 \\ 
                {\dot{{x}}}_{{p}} \end{array}
                \right]{\ }
    \end{split} \label{AikLik}
\end{align}


\subsection{Model Discretization}
\noindent After linearization, discretization is the next critical step in the model formulation. In this paper a Zero Order Hold (ZOH) is applied to discretize the model using matrix exponential method \cite{d:94}. In addition, similar to \cite{d:42}, prediction horizon is discretized with variable time steps approach. To meet the real time computation requirement, the number of discretization steps is limited to $N_p = 33$ in accordance with the computational power. Hence, the step size depends on the tradeoff between modeling and prediction accuracy, and length of prediction horizon. Obviously larger prediction horizon is preferred, but it results into larger step size reducing prediction accuracy. 
In this study, two time steps were  used namely, short time-step ($T_{ss}=30 \ ms$) and long time-step ($T_{ls} = 200 \ ms$). $27$ short time steps, i.e. $N_{{ss}}{=27}$ and $6$  long time steps i.e. $N_{{ls}}{=6}$ are selected to provide satisfactory path tracking and vehicle stabilization performance. The total of $N_{p}{=} {N_{{ss}}}{+}{N_{{ls}}}{=33}$ steps results in a time horizon of  ${2.1\ s}$. 
The number of short time steps selected are more than the control horizon steps $N_c =10$ selected for the proposed system, thereby, providing the controller a sufficient control over the dynamics involved. Moreover, the  the designed  control system's execution rate is set at  $30 \ ms$, and the short time steps at this rate provides required fidelity for vehicle dynamics modeling and fast convergence in the case of rear tire force successive linearization.  In our evaluation, the computation time for the convex optimization was approximately recorded as $10 \ ms$ on a single core of CPU E5-1650 @ 3.5 GHz. Therefore,  the $30 \ ms$ execution rate of designed control system, provides more time for the developed system to evaluate multiple admissible convex tube spaces as discussed in (\ref{CollCon})  \\
Equation (\ref{LinearTVModel})  represents the model after discretization with the ZOH and this Linear Time Varying (LTV) model is utilized for prediction in RMPC based control system in this paper.

\begin{align}
    {X}_{{i+1|k}\ }{=}
        \left[
            \begin{array}{cc}
            {{A}}^{{i|k}}_{{stab,d}} & {\overline{{A}}}^{{i|k}}_{{aug,d}} \\ 
            {{A}}^{{i|k}}_{{aug,d}} & {\textbf{1}} \end{array}
        \right]{\ X}_{i|k}{+}
        \left
            [ \begin{array}{c}
            {{B}}^{{i|k}}_{{aug,d}} \\ 
            0 \end{array}
        \right]{u}_{i|k}   \label{LinearTVModel} \\ \nonumber {+}
        \left[ 
            \begin{array}{c}
            {{L}}^{{i|k}}_{{stab,d}} \\  
            {{L}}^{{i|k}}_{{aug,d}} \end{array}
            \right] {+\ w_{i|k}} , \ \ {for\ i\ =\ 0\ \dots \ N_{p}}
\end{align}

\section{Robust Model Predictive Controller (RMPC)}
\subsection{Control Structure and feedback Gains}

\noindent The control input structure for the robust MPC framework in this study is defined by:

\begin{equation}
    {{u}}^{{*}}_{0|k}{=}\left[{{K}}^{{0|k}}_{{aug}}{\ 0}\right]{{X}}_{0|k}{+}{{c}}_{0|k}
    \label{controlinputstruct}
\end{equation}

\noindent where, ${{u}}^{*}_{0|k}{\ }$represents the control input computed at controller execution step ${k}$, ${{K}^{0|k}}_{{aug}}$ is the infinite horizon Linear Quadratic Regulator (LQR) optimal feedback gain for the pair (${(}{{A}}^{{0|k}}_{{stab}},{{B}}^{{0|k}}_{{stab}}$). The ${{c}}_{0|k}$ is the first output of the open loop MPC optimization. The LQR optimal gain ${{K}^{i|k}}_{{aug}}$, needs to be obtained for each prediction step ${i}$ using the method proposed in  \cite{d:95} due to the time-varying nature of the linear model described by (${(}{{A}}^{{0|k}}_{{stab}},{{B}}^{{0|k}}_{{stab}}$). 
In order to meet the real time constraint, optimal ${{K}^{i|k}}_{{aug}}$ are computed only for first $N_c$ short time steps and 1 long time step as shown in the equations below. 


\noindent 

\begin{align}
    {{u}}_{{i|k}} &{=}\left[{{K}}^{{i|k}}_{{aug}} \ 0 \right]{{X}}_{{i|k}}{+}{{c}}_{{i|k}}{,\ \ \forall \; i=0\dots .\ }{{N}}_{{c}} \label{controlpolicync}\\
    {{u}}_{{i|k}} &{=}
        \left[{{K}}^{{N_c|k}}_{{aug}} \ 0 \right]{{X}}_{{i|k}}{{+c}}_{{{{N}}_{{c}}}{|}{{k}}}{,\ \forall \; i=}{{N}}_{{c}}{\ +1,\dots }{,}{{N}}_{ss}-1 \label{controlpolicyntls}\\
    {{u}}_{{i|k}}&{=} 
        \left[{{K}}^{{N_{ss}|k}}_{{aug}} \ 0 \right]{{X}}_{{i|k}}{+}{{c}}_{{{Nc}}{|}{{k}}}{,\ \forall \; i=}{{N}}_{{ss}}{,\dots  , N_p-1} \label{controlpolicynp}
\end{align}

For the optimal LQR at each step, there are two (possibly) conflicting objectives, namely the stability of the closed loop system in presence of disturbances and physical limits on steering angle. The typical trade-off between these objectives is handled by appropriately choosing weights on each objective. 

\subsection{Stabilization Constraints}

\noindent Each vehicle has limitations in its handling capabilities due to existence of limitations in the grip of the tire. These limitations come forth whenever the vehicle requires forces from the nonlinear regions of the tire force curves during a maneuver. To operate the vehicle at these limits without allowing it to destabilize is a high skill task. In the case of autonomous vehicles, the controller can be expected to have an envelope of handling limits. Various ways have been proposed in past to consider vehicle stability limits. 

For example, \cite{d:53} and \cite{d:40} use states phase plots with side-slip angle for constant steering input to find the unstable regions. 
Based on the tire maximum force generation and understanding of the equilibrium point trends for the range of steering and braking inputs, \cite{d:40} introduced an invariant stable handling envelope for the vehicle. With this knowledge of a safe handling envelope , \cite{d:41} further showed that unlike Electronic Stability Control (ESC), which only acts when the vehicle is in the unstable region, the approach for stabilizing the autonomous vehicle should prevent it from entering the unstable region. This is the approach we have followed in the proposed control. 

Similar to \cite{d:40}, we bound our vehicle's lateral velocity ${\dot{{y}}}_{{p}}$ and ${r}$ (vehicle yaw rate at CG) by using the maximum steady-state force generation of the tires. 
In order to ensure the stability of the vehicle at a given longitudinal speed, the rear tire slip angle is constrained to the linear region of tire force curve. In this study, the rear tire saturation slip angle (${{\alpha }}_{{r}}{)}$ is converted into a bound on $\dot{{y\ }}$ and by using (13), it is translated then to a bound on ${\ }{\dot{{y}}}_{{p}}$. It is formally expressed for a given longitudinal speed ($\dot{x}_p$) as 
\begin{equation}
    {\dot{{y}}}_{{p,max}\ }{=}{\ \dot{{x}}}_{{p}}{{\alpha}}_{{r,sat}}{+}\left({p+b}\right){r\ } 
    \label{stabconlatvelmax}
\end{equation}

\noindent where, ${\dot{{x}}}_{{p}}$ represents the longitudinal velocity at CP, as shown in fig. (\ref{V_Model}), and $\alpha_{r,sat}$ is the rear tire saturation angle defined in the tire model (\ref{brushtiremodel}). \\

\begin{figure}\vspace{-0.5cm}
   \centering
     \begin{subfigure}[t]{0.45\textwidth}
        \includegraphics[height=2in,width=\linewidth]{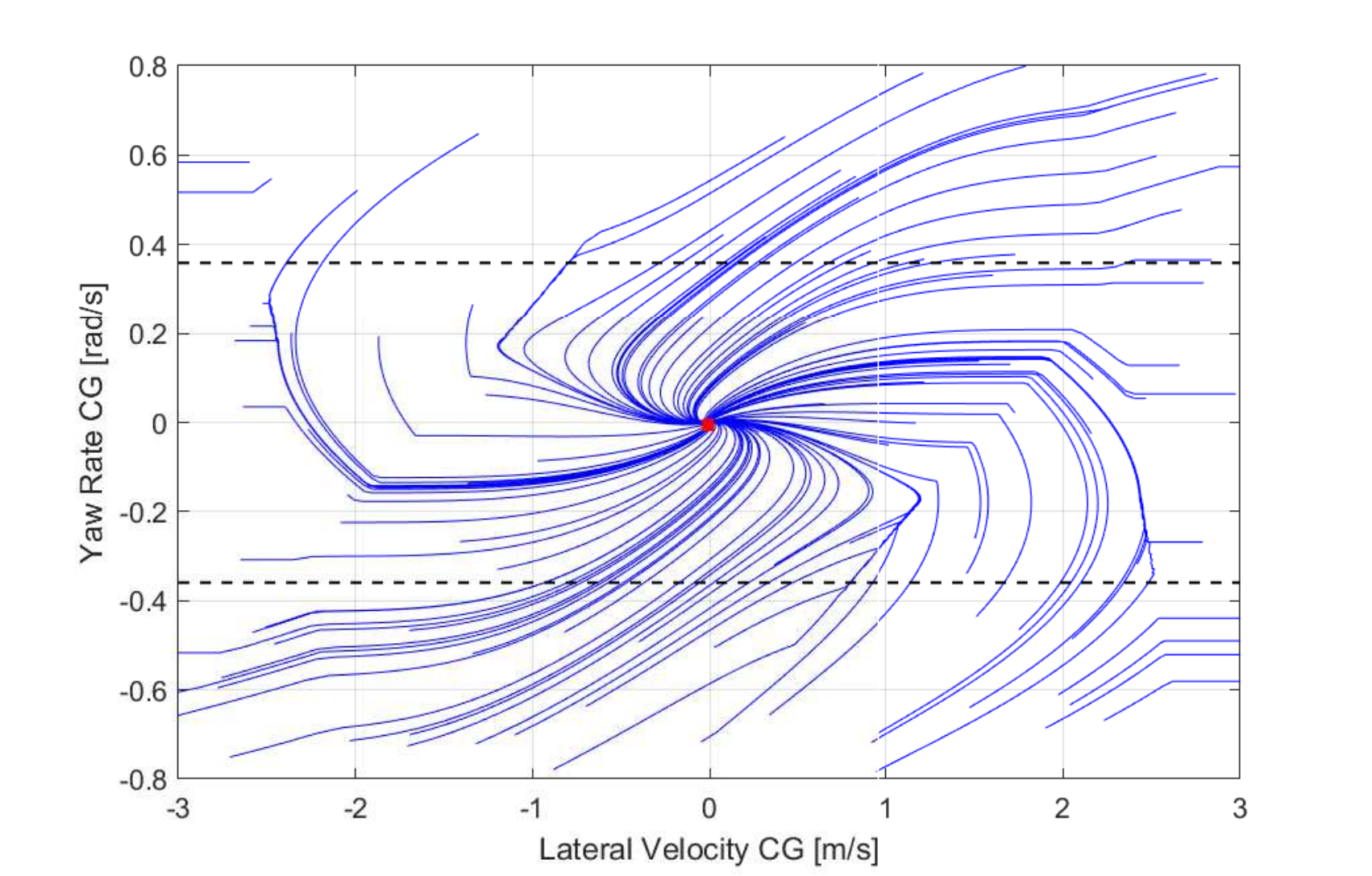}
        \caption{}
    \end{subfigure}
    \begin{subfigure}[t]{0.45\textwidth}
        \includegraphics[height=2in,width=\textwidth]{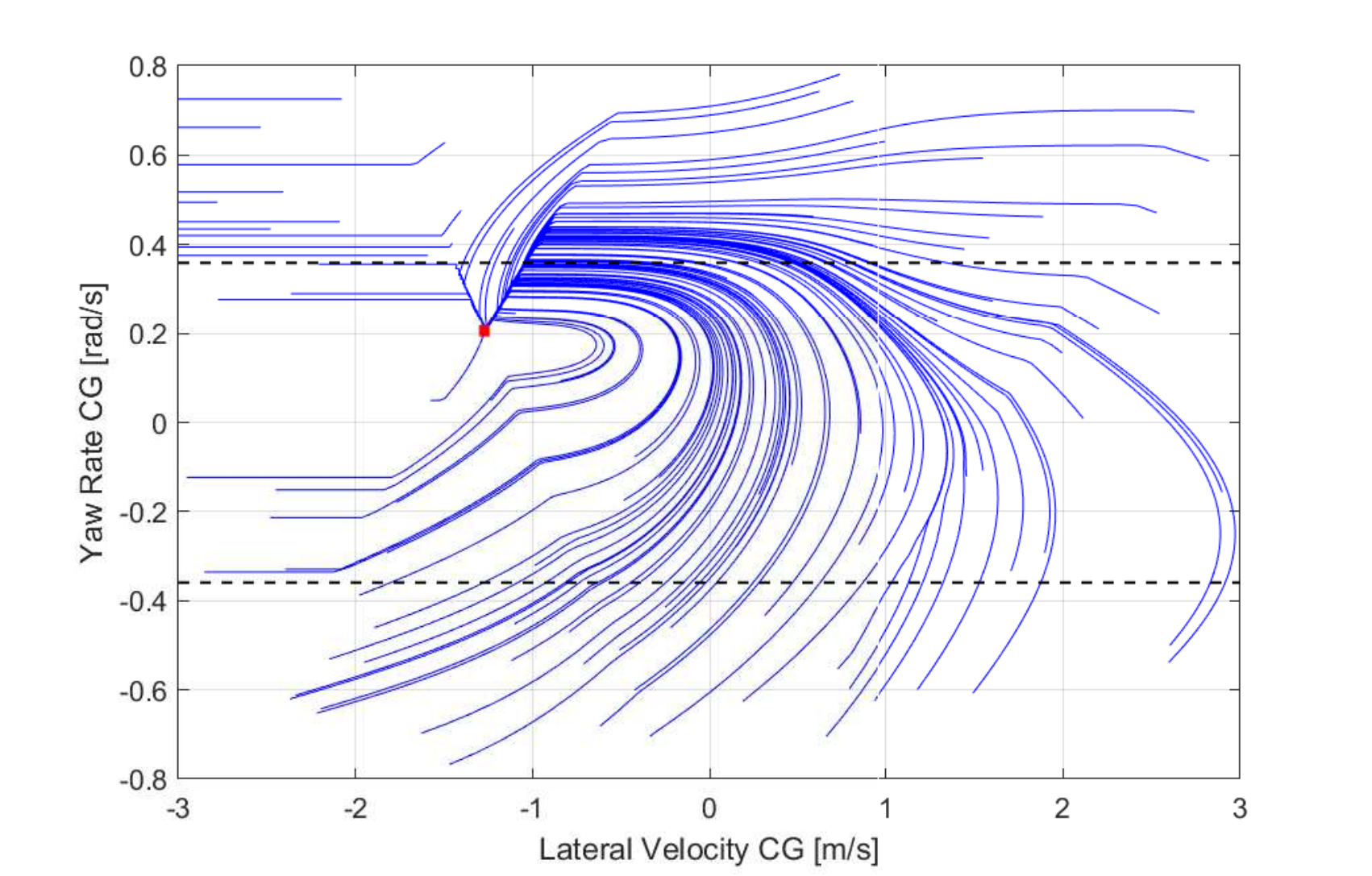}
        \caption{}
    \end{subfigure}

    \caption{The phase portrait of a neutral steering vehicle at $\dot{x}_p$=15 m/s, $\mu=0.55$, static front axle load $Fz_f =7.4$ kN and rear axle load $Fz_r = 4.93$ kN, and (a) $0^{\circ}$ and (b) $5^{\circ}$. The thicker dotted lines represents, the maximum-yaw rate computed in (\ref{stabconrmax}), examples of the trajectories as thin lines and stable equilibrium as heavy dot marker.}\label{figphaseportrait}
\end{figure}

The other constraint is on the yaw rate ($r$).
The bounds as stated in (\ref{stabconrmax}) is based upon the steady state yaw rate condition. If the maximum force at front axle $F_{yf,max}$ is less than the rear maximum force $F_{yr,max}$, the maximum yaw rate is then governed by the front axle, $F_{yf,max}$. 
In this work, constant longitudinal speed has been assumed and therefore, the vehicle is assumed to have a neutral steer. The neutral steer assumes simultaneous tyre saturation of the front and rear axles. Also the nominal forces on the rear and front tires are approximated by the static forces $F_{zf}$ and $F_{zr}$, respectively. This assumption hold true due to no brake actuation utilized by the controller. This transforms the minimum condition in (\ref{stabconrmax}) to a single value of $\mu g/{\dot{x}_p}$, where $\mu$ is the tire road friction coefficient and $g$ denotes the acceleration due to gravity. The phase plane plots of yaw rate v/s slip angle for a neutral steer has been shown in \cite{d:58}. These plots shows the trajectories for a range of yaw rate and slip angle combination, with a fixed steering and longitudinal speed. For the neutral dynamics \cite{d:53} \cite{d:58}, showed that there is a single, stable equilibrium point and two saddle points for steering angles upto $10\degree$, at a speed of $10$ m/s. Fig. (\ref{figphaseportrait}) shows the progression of the neutral steer dynamics for the two steering angles, at a speed of $ 15m/s$, in the phase plane plots similar to \cite{d:58}. Due to presence of saddle equilibrium in the region of high yaw rate(greater than the dotted lines) and high side slip angles, the trajectories diverge and for the initial conditions starting within the yaw rate bound converges to the stable equilibrium point. This shows that the steady state yaw rate bifurcates the phase plane plot into stable region and unstable region for different steering angles at a fixed longitudinal speed. Therefore, the bound on yaw rate, described by (\ref{stabconrmax}), is selected as the second set of stability constraints in this study.

\begin{align}
    {{r}}_{{max}\ }&{=}{{min} \{{\ }\frac{{{F}}_{{yf,max}}{(1+\ }{{a}}/{{b}}{)}}{{m}{\dot{{x}}}_{{p}}}\ }{,\ }\frac{{{F}}_{{yr,max}}{(1+\ }{{b}}/{{a}}{)}}{{m}{\dot{{x}}}_{{p}}}{\ }{\}}{\ \ } \label{stabconrmax}\\
    \nonumber &=\frac{\mu g}{\dot{x}_p} \ 
\end{align}

\begin{figure}
    \centering
    \includegraphics[width=0.5\textwidth, height=2.5in]{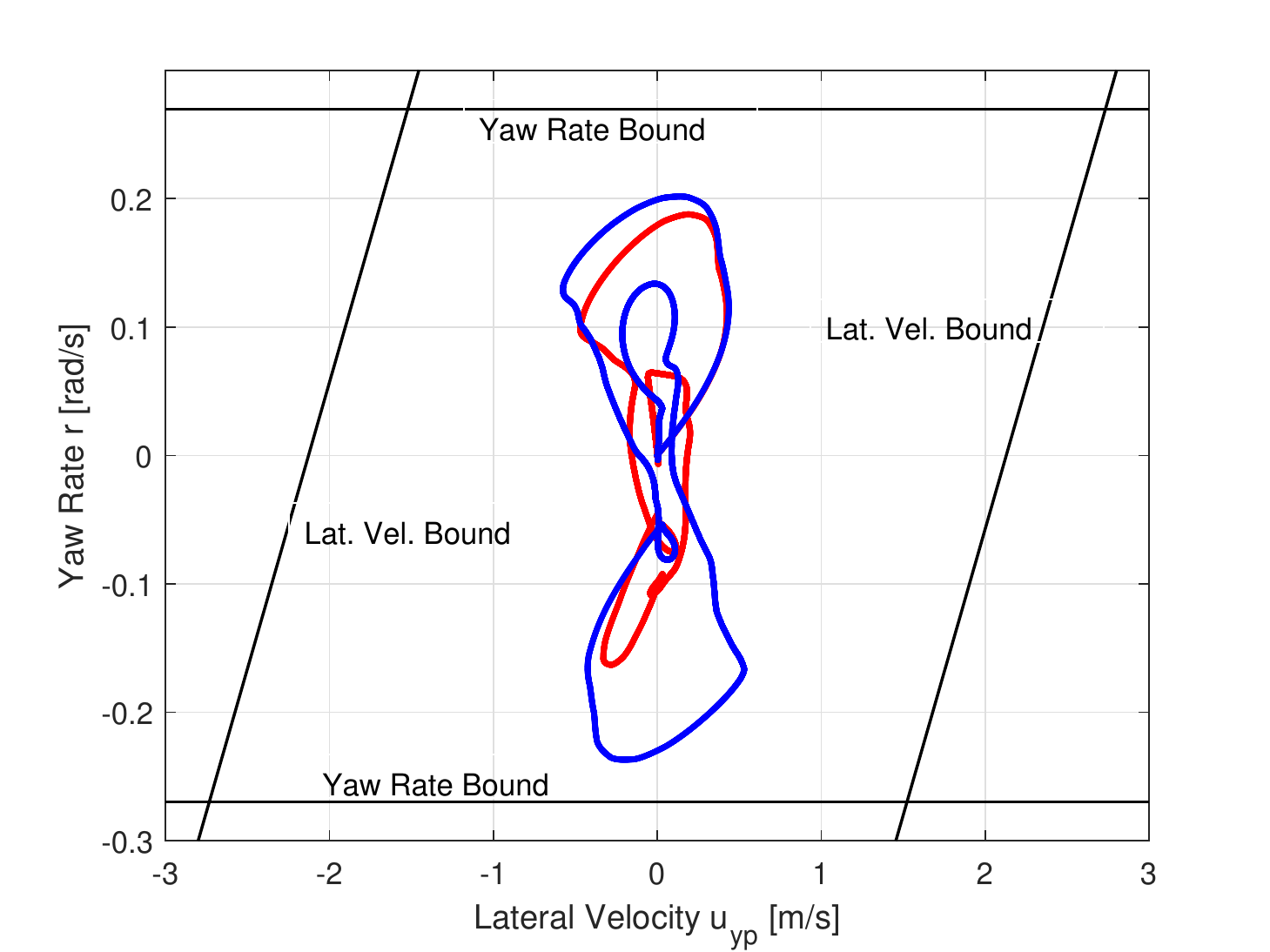}
    \caption{Stability Envelope. A closed bounded set represented by the yaw rate and lateral velocity (CP) bounds in a yaw rate v/s lateral velocity phase plane. The blue and red are examples of two safe trajectories of the vehicle that adhered to the safety envelope.}
    \label{figStabEnv}
\end{figure}

The bound on CP lateral velocity in (\ref{stabconlatvelmax}) and the yaw rate in (\ref{stabconrmax}), creates a closed bound set for the phase plane of lateral velocity (CP) and yaw rate, as shown in fig. (\ref{figStabEnv}). Vehicle stability is guaranteed for all states in the envelope formed by (30) and (31). The invariance of the set has been proved in the \cite{d:40}. Exceeding not necessarily cause the instability of the vehicle, as it was shown in \cite{d:42} \cite{d:43}, however, for the states outside the envelope, the availability of a steering control input to move the vehicle closer to the boundary in the next time step cannot be guaranteed.\\
 
 In this work, during optimization the yaw rate constraint is penalized more than on the lateral velocity due to designed control system planning for trajectories that utilizes the maximum rate of change of force inputs for obstacle avoidance.  

The stabilization constraint for the controller is organized in the form expressed as: 
\begin{equation}
    {{E}}_{{stab}}{{X}}_{{i|k}}{\ }{\le }{\ }{{G}}_{{stab}}
    \label{stabconstraint}
\end{equation}

\noindent where the ${{E}}_{{stab}}{\in }{\mathbb{R}}^{{l\times n}}$ and ${{G}}_{{stab}}{\in }{\mathbb{R}}^{{l\times 1}}$.

\subsection{Collision Constraints}\label{CollCon}

\begin{figure}
   \centering
     \begin{subfigure}{0.45\textwidth}
        \includegraphics[width=0.97\textwidth]{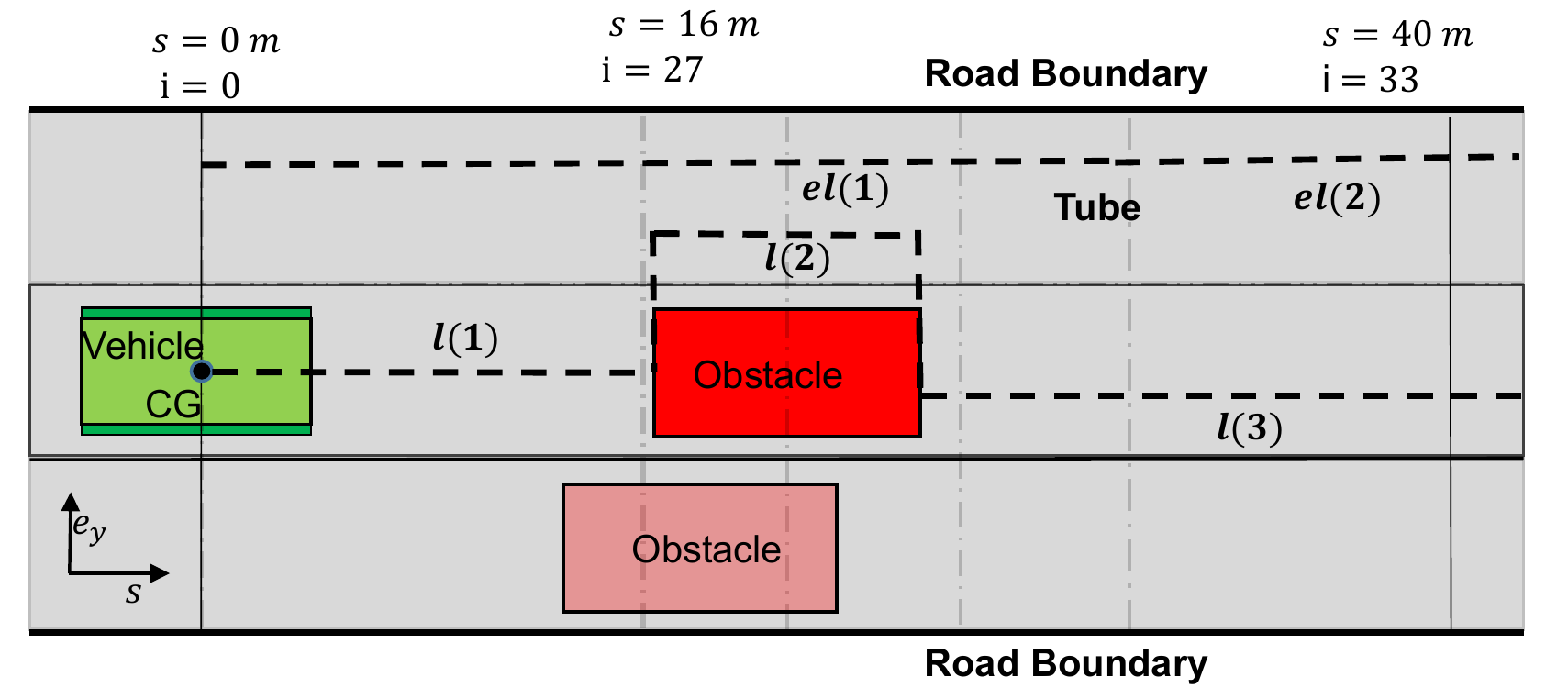}
        \caption{ Active Constraints. $el(1)$, $el(2)$, $l(1)$, $l(2)$ and $l(3)$ represent active constraints in the prediction horizon. Tube represented by combining the active constraints.}
           \label{figCollconActiveCon}
    \end{subfigure}
    \par\bigskip
   \begin{subfigure}{0.45\textwidth}
       \includegraphics[width=0.97\textwidth]{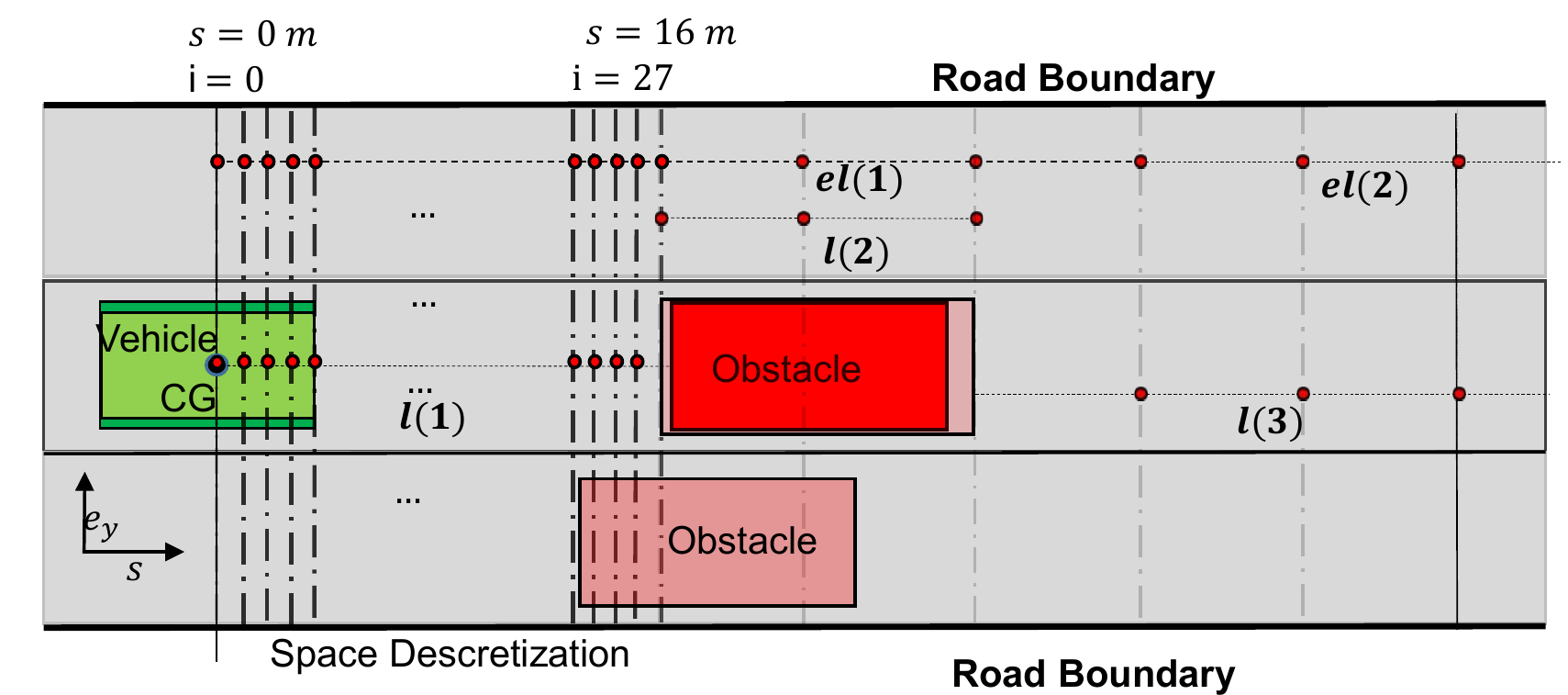}
           \caption{ Time Varying Space Constraints. The heavy red dots at a fixed path distance, forms the bound on the lateral error $e_y$. The transparent red box around the obstacle represents the stretched obstacle lengths. Dash-dot lines illustrate Prediction horizon space discretization with both time steps, $T_{ls}$ long, and $T_{ss}$ short.}
        \label{figSpaceDiscretization}
    \end{subfigure}
      \par\bigskip
  \begin{subfigure}{0.45\textwidth}
      \includegraphics[width=0.97\textwidth]{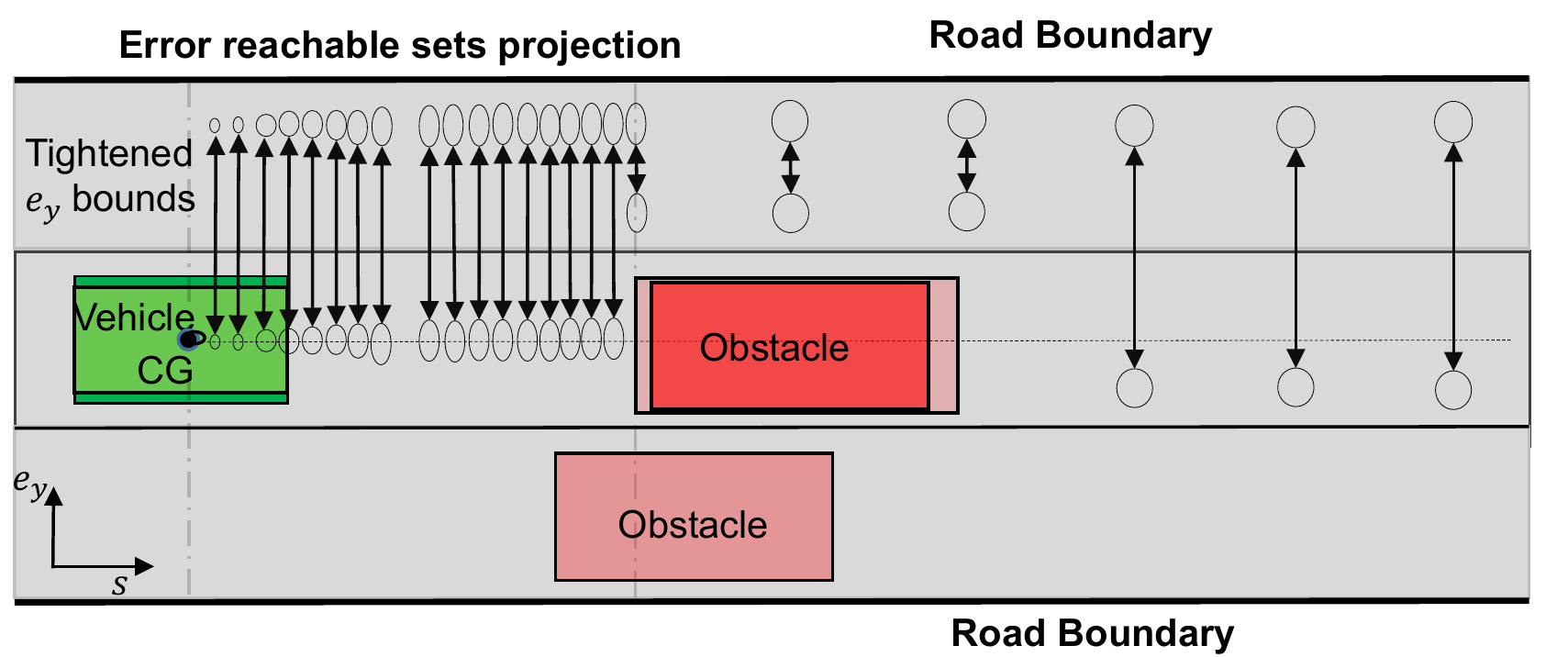}
     \caption{ Robust Collision Constraints. The ellipsoids represent the error reachable set projection on the time varying space constraints. The double arrow line at a prediction step $i$ represents the tightened bound for the lateral error state $e_{y}^{i|k}$. Dashed line represents the desired path.}
        \label{figRobustCon}
    \end{subfigure}
    \caption{Illustration of robust collision constraints development process at a controller execution step $k$. Vehicle assumed to have longitudinal speed $\dot{x}_p$ = $20$ m/s  }\label{figCollCon}
\end{figure}

\noindent In order to ensure that the planned trajectories avoid the obstacles and stay within the edges of the road, the collision constraints are defined for each control execution step. Obstacles and road edges combined together form an admissible region (search space) for the path planning objective around the ego vehicle. In real-world scenarios, each obstacle provides two options to pass an obstacle, either from the right or from the left (if the space permits). In this section, only left pass constraint computation has been explained. It is assumed the decision to take left or right can either be an output of the higher level decision layer or could be an outcome of simultaneously solving the optimization for both set of admissible regions and selecting the direction based on the one with the reduced cost. Therefore, without loss of generality, we have shown a maneuver from left of the obstacle in fig. (\ref{figCollCon}).

 The admissible region can be defined by the constraints called active constraints. In this study we assume that the perception layer (Camera, RADAR etc. along with detection algorithm and coordinate transformations) is capable of generating the locations of obstacles and road edges in the path reference frame [$e_y$,$s_d$]. These locations are then expressed in form of linear inequalities on path error states as shown in fig. (\ref{figCollconActiveCon}). 
 
 The introduction of error states simplifies the constraint process, as the stability constraints are defined using the states represented in body centered frame of reference, whereas collision constraints on the error states. 

\begin{align}
    {{E}}^{{j}}_{{coll}}{{X}}_{{i|k}}{\ }{\le }{\ }{{G}}^{{j}}_{{coll}}{\ ,\ j=1\dots l} 
    \label{TimeVaryingCollCon}
\end{align}

\noindent where $l$ denotes number of active constraints for the states at instant $k$, that can be combined to form a tube or a search space around the ego vehicle as shown in fig. (\ref{figCollconActiveCon}). The tube defined is then discretized using the path distance estimation ${s_d}_{i|k}$ for each prediction step. The path distance is estimated using the discretized distance equation in (\ref{LinearTVModel}) and the information of time steps $T_{ss}$ and $T_{ls}$. The transformed tube is represented as 

\begin{equation}
    {{E}}^{{i|k}}_{{coll}}{{X}}_{{i|k}}{\ }{\le }{\ }{{G}}^{{i|k}}_{{coll}} 
    \label{TVCollConSimplified}
\end{equation}  
\noindent where ${{E}}^{{i|k}}_{{coll}}{\in}{\mathbb{R}}^{q \times n}$ and ${{G}}^{{i|k}}_{{coll}}{\in}{\mathbb{R}}^{q }$ consists of constraints on the normal path error state ${e_y}_{i|k}$, defining the gap in the search space for the $i^{th}$ prediction step as shown in fig. (\ref{figSpaceDiscretization}). It should be be noted that before the discretization step the obstacle lengths are stretched, illustrated in fig. (\ref{figSpaceDiscretization}). The example illustrated in fig. (\ref{figCollCon}) is shown for the obstacles that are stationary and it is assumed that no information of their intended trajectories are available. 
However, if the prediction is available, the tube could be adapted due to its time varying nature.

Now that the discretized admissible region in established, the next step is to include projection of the errors induced by the additive disturbance by tightening the constraints in (\ref{TVCollConSimplified}). 
The equation (\ref{TVCollConSimplified}) can be expressed with constraints tightening as:

\begin{align}
     {{E}}^{{i|k}}_{{coll}}
        \left({{s}}_{{i|k}}{+}{{e}}_{{i|k}}\right){\ \le }{\ {G}}^{{i|k}}_{{coll}}{,\ \ \ \ \ \ \ \ i=0,1,\ .\ \dots }{{N}}_{{p}} 
\label{TVCollConse}
\end{align}

\noindent where ${{s}}_{{i|k}}$ and denote $e_{i|k}$ the nominal state and error in the state induced due to disturbance at the prediction step $i$, respectively. The time varying constraint (\ref{TVCollConse}) can be stated as 

\begin{align}
    {{E}}^{{i|k}}_{{coll}}{{s}}_{{i|k}\ }{\le }{\ {G}}^{{i|k}}_{{coll}}{-}{{h}}_{{i}}{,\ \ \ \ \ \ \ \ i=0,1,\ .\ \dots }{{N}}_{{p}}
\label{TVCollConEGh}    
\end{align}

\noindent  where ${{h}}_{{i}}$ is defined as

\begin{align}
    {{h}}_{{i}\ }{=\  }{\mathop{{max}}_{{{e}}_{{i|k}}{\in }{{S}}^{{i}}} {{E}}^{{i|k}}_{{coll}}{{e}}_{{i|k}}\ }{,\ \ \ \ \ \ \ \ i=1,2,\ \dots }{{N}}_{{p}} 
\label{hdef}
\end{align}

\noindent The ${{S}}^{{i}}$ in (\ref{hdef}) represents the reachable sets for the state error due to disturbance at each prediction step till step ${i}$. These reachable sets follow the constraint 

\begin{align}
    {{\phi }}^{{i}}{{S}}^{{i}}{ \oplus }{W}{\ \subseteq }{\ {S}}^{{i+1\ }}{\ }
\label{ReachableSets}    
\end{align}

\noindent where ${{\phi }}^{{i}}{=}{{A}}^{{i}}{+}{{B}}^{{i}}{{K}}^{{i}}$. The (\ref{hdef}) using the (\ref{ReachableSets}) constraint can be redefined as   

\begin{align}
    {{h}}_{{i}}{=}\sum^{{m=i-1}}_{{m=0}}{{\mathop{{max}}_{{{w}}_{{m|k\ }}{\in }{W}} \left[{{E}}^{{i|k}}_{{coll}}{.
    \left\{\prod^{{k=0}}_{{k=i-m-1}}{{{\phi }}^{{k}}}\right\}{.w}_{m|k}}\right]\ }}
    \label{hcomputation} \\
    \nonumber {\ ,\ i=1,2,\ \dots N_p}
\end{align}

 In equation (\ref{hcomputation}), computation of $h_i$ for $N_p$ steps require total number of $(2N_p+1)q$ linear programs, where $N_p$ denotes the prediction horizon and q denotes the number of inequalities present at step $i$. Now, due to discretization of space, required for formulation (34), reduced the number of linear inequalities to $q=2$ for each prediction step $i$, as shown in fig. (\ref{figSpaceDiscretization}) by the red dots. This reduction in $q$ resulted in limiting the number of linear programs required for $h_i$ to $(4N_p+2)$. The result shows that the number of linear programs to be solved are independent of any number of obstacles in the search space of the ego vehicle. This is a significant result from the point of keeping the additional computational load required for providing robustness in check. 

Note that in equation (\ref{hcomputation}), the $h_i$ are defined for the entire prediction horizon. In practice the MPC control considers only $N_c$ steps. This helps in avoiding the excessive reduction of the search space for prediction steps beyond the control horizon for which controller is incapable to plan. Also, in order to reduce computation time, the $h_i$ terms for $N_c+1$ to $N_p$ are considered constant, while the $h_i$ are explicitly computed for first $N_c$ steps as shown in the equation below.


\begin{align}
    {{h}}_{{i}} {=}\sum^{{m=i-1}}_{{m=0}}{{\mathop{{max}}_{{{w}}_{{m|k\ }}{\in }{W}} \left[{{E}}^{{i|k}}_{{coll}}{.
        \left\{\prod^{{k=0}}_{{k=i-m-1}}{{{\phi }}^{{k}}}\right\}{.w}_{{m|k}}} \right],\ }} \label{hcomputationsimplified}
        \\ \nonumber {\ i =1,2,\ \dots N_c}, \
    {{G}}^{{i|k}}_{{coll}}{{\ge }{h}}_{{i}}{\ge }\boldsymbol{{0}} \\ \nonumber
    {{h}}_{{i}} \ {=} \ {{h}}_{{{N}}_{{C}}}, \ \ \ \ \ \ \ \ \ i={{N}}_{{c}} {+1\dots \dots }{{N}}_{{p}}, \ \ {{G}}^{{i|k}}_{{coll}}{\ge }{{h}}_{{i}}{\ge }\boldsymbol{{0}}
\end{align}

 In (\ref{hcomputationsimplified}), the selected bounded disturbance set ${W}$ is a simple bound based on the expected modelling error and  state measurements error. It is a critical set and the next section (IV) in this paper covers the process of determining it in more detail.


We further simplified the constraints and implemented as a time varying bound on the lateral error, $e_y$, state, as shown in  fig. (\ref{figRobustCon}) and expressed as 

\begin{align}
    {\overline{e}}_{{y,\ k+i|k,min}}{\ }{\le }{\ }{{e}}_{{y,i|k}}{\ }{\le }{\ }{\overline{e}}_{{y,k+i|k,max}}{\ ,\ \ } 
\label{patherrcon}
\end{align}

The tube formed by the time dependent constraints in (\ref{patherrcon}) are not convex and thus, further requires convexification for the convex optimization. Similar to successive linearization concept that uses MPC's previous step optimization information, we use heading error angle state ($e_\phi$) from the last step to adjust the bounds in (\ref{patherrcon}). It is expressed as                                                                                                             
\noindent 

\begin{align}
    \begin{split}
        &{\overline{e}}_{{y,\ k+i|k,min}}{\ +\ f_{width}}
            ({e_{\phi }}_{{i|k-1}}{ }){\ \le }{\ {e}}_{{y,i|k}}{\  } \\ &{\overline{e}}_{{y,k+i|k,max}}{\ \ -\ f_{width}}({e_{\phi }}_{{i|k-1}}) {\ \ge}{\ }{{e}}_{{y,i|k}}{\ \ \ \ \ } 
     \end{split}
\label{patherrorconconvex}
\end{align}

\noindent where, the function ${f_{width}}\left({{\phi }}_{{i|k-1}}\right)$ approximates the vehicle's effective width \cite{d:65} based on the orientation information gleaned from the prediction step ${i}$ of last time instant $k-1$ optimization as the path changes. If $e_{\phi} = 0$, the${\ }{{f}}_{{width}}\left(e_{\phi }{=0}\right){=}\frac{{w}}{{2}}$, while at ${e_{\phi} }{=}{{\pi/2}}$, the approximate width is  ${{f}}_{{width}}\left({e_{\phi} }{=}{{\pi/2}}\right){=a}$, the distance from the center of gravity to the front axle. 

 With the time varying perception constraint defined in (\ref{patherrorconconvex}), the next section formally introduces the optimal problem solved at every time instant${\ k}$ for the designed control system and it will be followed by the section on results and discussions.

\subsection{Optimization Problem}
\noindent In this section we formulate the path tracking and obstacle avoidance online RMPC problem for an autonomous vehicle. RMPC, at each sampling time instant $k$, computes an optimal input sequence or specifically, a lateral force input sequence, by solving a constrained finite horizon optimal control problem. This computed optimal input sequence and its corresponding optimized state trajectory are kept as the nominal input and state trajectories, $\mathbf{c}$ and $\mathbf{s}$ respectively. Using first input of the computed sequence $c_{0|k}$ and the control structure in (\ref{controlinputstruct}), the augmented control input $u_{0|k}$ is calculated, which is then applied to the vehicle at current step. At the next time instant, the constrained optimization is setup again and solved using a new set of measurements.

\begin{align} 
    \begin{split}
        {{s}}_{{i|k+1}\ }{=}{\ {(A}}^{{i|k}}{+}{{B}}^{{i|k}}{{K}}^{{i|k}}{)}{{s}}_{{i|k}}{+}{{B}}^{{i|k}}{{c}}_{{i|k}}{+}{{L}}^{{i|k}} \\ {=}{\ {f}}^{{i|k}}{(}{{s}}_{{i|k}},{{c}}_{{i|k}}{)}
    \end{split}
    \label{optimvehmodel}
\end{align}

In this study, the cost function of the optimal control problem is designed to mediate between objectives defined by trajectory re-planning, stability constraints, and the collision constraints. The augmented input is the front tire force  calculated based on the discretized model in (45) and converted to a steering command using the tire model in (\ref{brushtiremodel}) and the relationship in (\ref{slipangles}). The nominal optimization problem for the optimal control problem is represented as

\begin{align}
    \begin{split}
        &\mathop{{min}}_{\boldsymbol{{c}},{\epsilon_{stab}},{\epsilon_{coll}}} \sum^{{{N}}_{{p}}{-}{1}}_{{i=0}}{{
            \left\|{{s}}_{{i|k }}\right\|}^{{2}}_{{Q}}}{+}{{\left\|{{c}}_{{i|k }}\right\|}^{2}_{{R}}}{+}{{
            \left\|{\Delta}{{c}}_{{i|k }}\right\|}^{{2}}_{{S}}}
        \\ 
        & \ \ \ \ \ \ \ \ \ \ \ \ \ \ \ \ \ \ \  {+}{{
            \left\|{\epsilon_{stab}}\right\|}^{{2}}_{{\lambda_{stab}}}}{+}{{
            \left\|{\epsilon_{coll}}\right\|}^{{2}}_{{\lambda_{coll}}}}
    \end{split} \tag{45a} \label{optproba}\\ 
    \begin{split}
        &{{s}}_{{i|k+1}}{=}{{f}}^{{i|k}}{(}{{s}}_{{i|k}},{{c}}_{{i|k}}{)}   {, \ i=0,1,\dots \dots   ,}{{N}}_{{p}}{-}{1}
    \end{split} \tag{45b} \label{optprobb}\\ 
     \begin{split}
        &{\overline{{e}}}_{{y,i|k,min}}{-}{\epsilon_{coll}}{\le }{\ }{H}{{s}}_{{i|k}}{\le }{\ }{\overline{{e}}}_{{y,i|k,max}}{+}{\epsilon_{coll} } \\ & \ \ \ \ \ \ \ \ \ \ \ \ \ \ \ \ \ \ \ \ \ \ \ \ \ \ \ \ \ \ \ \ \ \ \   {,\ i=0,1,\dots ..,\ Np} 
     \end{split} \tag{45c} \label{optprobc}\\
    \begin{split}
         &{{E}}^{{k}}_{{stab}}{{s}}_{{i|k}}{\ }{\le }{{G}}^{{k}}_{{stab}} {+}{\epsilon_{stab}} \\ & \ \ \ \ \ \ \ \ \ \ \ \ \ \ \ \ \ \ \ \ \ \ \ \ \ \ \ \ \ \ \ \ \ \ {,\ i=0,1,\dots ..,\ Np} 
    \end{split} \tag{45d} \label{optprobd}\\
        &\epsilon_{stab}\ge \mathbf{0},\; \epsilon_{coll}\ge 0 \tag{45e} \label{optprobe}\\
        &{{c}}_{{i|k}}{=}{\Delta }{{c}}_{{i|k}}{+}{{c}}_{{i-1|k}} \tag{45f} \label{optprobf}\\ 
        &{{c}}_{{i|k}}{\in }{{\mathbb{U} }}{,\ \ \ \ \ \ \ \ \ \ \ \ i=0,1,\dots ..,}{{N}}_{{c}}{-}{1} \tag{45g} \label{optprobg} \\ 
        &{{\Delta }{c}}_{{i|k}}{\in }{\Delta }{\mathbb{U} }{,\ \ \ \ \ \ \ i=0,1,}{\dots }{..,}{{N}}_{{c}}{-}{1} \tag{45h} \label{optprobh}\\ 
        &{\Delta }{c}_{{i|k}}{=0 \ \ \ \ \ \ \ \ \ \ i=}{{N}}_{{c}}{,\dots ..,}{{N}}_{{p}} \tag{45i} \label{optprobi}
\end{align}

 \noindent 
where ${{s}}_{{i|k}}$ denotes the predicted nominal state at prediction time step ${i}$ obtained by applying the control sequence ${\boldsymbol{{c}}}_{\boldsymbol{{k}}}$ ${=}{\{}{{c}}_{{0|k}}{,\dots ,\ }{{c}}_{{{N}}_{{c}}{-}{1|k}}\}$ to the model in (\ref{optprobb}) with ${{s}}_{{0|k}}{=x_k}$. 
The $s_{ref,i|k}$ represents the reference path of the vehicleThe collision constraints (\ref{patherrorconconvex}) have been imposed as soft constraints, by introducing the slack variable ${\epsilon_{coll}}$ and ${\epsilon_{stab}}$ in (\ref{optprobc}) and (\ref{optprobd}) respectively. 
The collision avoidance and stabilization are sometimes conflicting objectives. Therefore, the weights ${\lambda_{stab}}$ and ${{\lambda_{coll}}}$, for the stability and collision constraints violation ensure problem feasibility, and their values encode the tradeoff between the two objectives. The robust collision constraints  by including modeling errors proactively, take care of the search space constraints and therefore, in the optimization, the stability constraint is enforced over the collision constraint. The weights $0 \prec Q\in \Re^5\times \Re^5$, $0<R \in \Re$ and $0<S\in \Re$ are penalises on the state tracking error, RMPC control action, and the rate of change of nominal the control action respectively.

It should be noted that the dynamic feasibility of the RMPC is difficult to guarantee under all the possible scenarios. This is due to the nonlinearities of the system and the changing search space at every time instant. However, proposed approach reduces the need for conservative fixed constraints (e.g. \cite{d:70}) by dynamically updating the limits.

\section{Simulation Study Results}

\begin{figure*}[tb]
    \includegraphics[width=\textwidth, height=10cm]{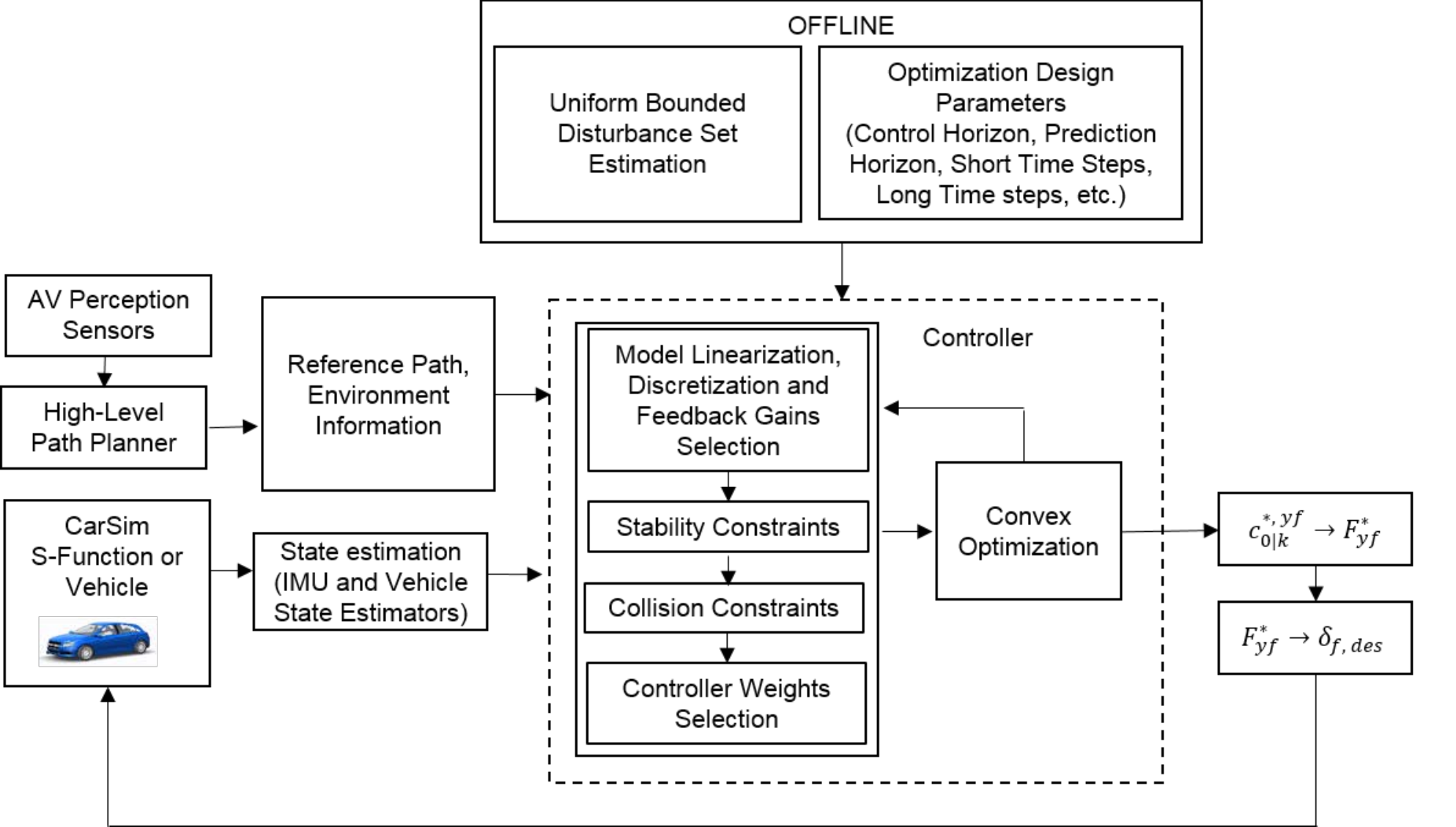}
    \caption{\label{figSimulation-Setup} The block diagram of robust model predictive control for obstacle avoidance and path tracking system of autonomous vehicle}
\end{figure*}

\noindent In this section, the designed control system is tested for specific test cases that are provided with a prescribed reference path and the initial longitudinal velocity as inputs. The performance of the designed system is compared  with a Linear Time Varying  non-robust Deterministic MPC (DMPC) controller equipped with successive linearization of the rear tire forces in order to understand the advantages of robust control design. Three different test cases are presented to demonstrate the system's performance envelope. Simulation is performed using CarSim linked with Simulink software and B-Class model is chosen. The vehicle parameters are shown in Table \ref{vehicleparams}. By linking CarSim to Simulink, the full vehicle model with high DOF and system's non-linearities is linked to the designed controller in Simulink. 
The majority of the researchers use software simulation for testing and validating their designed vehicle control systems \cite{d:81,d:82, d:83, d:84, d:85, d:86, d:87,d:88,d:89,d:90}.  The proposed system's block diagram with CarSim model-in--the-loop is shown in the fig. (\ref{figSimulation-Setup}). The convex optimization problem in (\ref{optproba}-\ref{optprobi}) and linear program in (\ref{hcomputationsimplified}) are solved using  CVXGEN solvers \cite{d:80}. As shown in fig. (\ref{V_Model}), the augmented force input computed by the MPC is converted into a steering command using the tire model discussed in (\ref{brushtiremodel}). The Table \ref{parametersandweights} lays out the weights for the convex optimization. The weights set in the controller prioritise the yaw rate bound enforcement, $\lambda_{stab}[3,4]$, over the collision constraint ${\lambda_{coll}}$.  Table \ref{ComputeResults} shows the computation time to optimize the resulting optimization problem (\ref{optproba}-\ref{optprobi}) on a single core of CPU E5-1650 @ 3.5 GHz \cite{d:80}.  

\begin{table}
    \begin{center}
        \begin{tabular}{l  c  c  c}

            \multicolumn{1}{ l}{Parameter} & \multicolumn{1}{c }{Symbol}  & \multicolumn{1}{c }{Value}& \multicolumn{1}{c }{Units}\\

            \hline
            Vehicle Mass & $m$ & 1260 &$kg$ \\
            
            Yaw Moment of Inertia & $I_{zz}$ & 1343.1 & $kg/{m^2}$\\
            
            Wheelbase & $L$ & 2.6 & $m$\\
            
            Front Axle- CG Distance & $a$ & 1.04 &$m$ \\
            
            Rear Axle- CG Distance & $b$ & 1.56 &$m$ \\
            
            Vehicle Width & $wd$ & 1.695 &$m$ \\
            
            Cornering stiffness of front tire & $C_{\alpha_f}$ & 51650 & $N/rad$ \\
            
            Cornering stiffness of rear tire & $C_{\alpha_r}$ & 38160 & $N/rad$ \\
            
            Normal force front tire &$Fz_f$ &2704.4 &$N$ \\
            
            Normal force rear tire &$Fz_r$ &2704.4 &$N$ \\
            
            Road surface friction coefficient & $\mu$ & 0.55 & {}\\
            
            Radius of the tire & $R_{tire}$ & 0.3 & $m$\\
            
            Height of CG & $h$ & 0.54 & $m$ \\
            \hline
        \end{tabular}
    \end{center}
    \caption{\label{B Vehcile}B-Class CarSim Vehicle Parameters } 
\label{vehicleparams}    
\end{table}

\begin{table}
    \begin{center}
            \begin{tabular}{m{3cm}m{0.5cm}m{2cm}p{0.25cm}}
        
                \multicolumn{1}{l}{Parameter Description} & \multicolumn{1}{c}{Symbol}& \multicolumn{1}{c}{Value } &\multicolumn{1}{c}{Units}\\
                \multicolumn{1}{c}{} & \multicolumn{1}{c}{}& \multicolumn{1}{c}{(Avoidance,Tracking)} &\multicolumn{1}{c}{}\\
                \hline 
        
                \textbf{Control System}\\
        
                Number of Timesteps&$N_p$&33, 33&{} \\
                Controller Timestep & $T_s$ & 0.03, 0.03 &$s$\\
                Short Timestep size  & $T_{ss}$ & 0.03, .03 &$s$\\
                Long Timestep size & $T_{ls}$ & 0.2, 0.2& $s$ \\
                Control horizon step & $N_c$ & 10, 10 & {} \\
                Timestep change  &$N_{ss}$ & 27, 27 & {}\\ \\

                \textbf{Weight Normalization}\\
        
                Max. expected lateral error  & $e_{y,max}$ & 3.8, 3.1 &$m$ \\
                Max. expected heading error  & $\Delta\phi_{max}$ & 0.2, 0.2 &$rad$\\
                Max. expected lateral velocity  & $u_{yp}$ & $0.2u_{xp}$, $0.2u_{xp}$ &$m/s$\\
                Max. expected yaw rate &$r_{max}$& $0.9\frac{\mu g}{u_{xp}}$,  $0.9\frac{\mu g}{u_{xp}}$ &  $rad/s$ \\
                Max. expected force input  &$\mathbb{U}_{max}$ & $\mu Fz_f$, $\mu Fz_f$&  $N$ \\
                Max. expected force input change &$\Delta\mathbb{U}_{max}$ & 12, 12 &  $kN/s$ \\
                 \\
        
                Vehicle states' weights & $Q_{diag}$ &[1,1,1,5,0], [1,5,1,10,0] & {} \\
                Force input  & $R$ & 4, 4 & {}\\
                Change in force input &$S$ & 4, 4 & {} \\
                Collision slack cost & $\lambda_{coll}$ & $10^5$, $10^5$ &{} \\
                Stability slack cost & $\lambda_{stab}$ & $10^2\times$ $[5,5,10^3,10^3]$, $10^4\times$ $[5,5,10^2,10^2]$ & {}
                \\
                \hline
            \end{tabular}
    \end{center}
\caption{\label{Control MPC}RMPC control system parameters and weights } 
\label{parametersandweights}
\end{table}

\begin{table}
    \begin{center}
        \begin{tabular}{l  c  c  c}
            \multicolumn{1}{ l}{Task} & \multicolumn{1}{c }{Time}  & \multicolumn{1}{c }{Time}& \multicolumn{1}{c }{Units}\\
            \multicolumn{1}{ l}{} & \multicolumn{1}{c }{(Avoidance)}  & \multicolumn{1}{c }{(Tracking)}& \multicolumn{1}{c }{}\\
            \hline
            Convex Optimization  & $9 \pm 1 $ & $7 \pm 2$ & $ms$ \\
            Robust Constraint Setup  & $4\pm 0.5$ & $4\pm0.5$ & $ms$\\

            \hline
        \end{tabular}
    \end{center}
\caption{\label{ComputeResults}Constrained optimization computation results} 
\end{table}

\noindent 
\subsection{Estimation of Disturbance Bound}

\noindent The RMPC requires estimation of the bounded disturbance set $W$ of the signal $w_{i|k}$ which represents an additive disturbance to the system model in (\ref{LinearTVModel}). We employ a process similar to \cite{d:45} to identify this set. We compare the one-step prediction of the model in (\ref{LinearTVModel}) with the measured vehicle states from the Carsim model. Multiple   one-step predictions were performed that mimic the situations encountered during evasive steering with particular emphasis on the mismatch of road friction coefficient errors. The example of one such test trial can be a double lane change maneuver with model assuming $\mu=0.7$ and Carsim model running at $\mu=0.5$. 
In addition to modeling error embedded with residual model uncertainty due to linearization, the estimation of the bound also includes the IMU measurement errors information from the EURO NCAP test protocol  \cite{d:96} measuring equipment requirements for lane support systems. After including all the aforementioned uncertainties, the set $W$ is defined by $W=\{w\in\Re^5| |w|<[0.2,0.14,0.0175,0.025,0.025] \}$.  

      

\subsection{Test Cases}
\noindent In this subsection, the test cases are defined for a curved road with a constant radius, $R =400$ m and tire-road friction coefficient $\mu \approx 0.55$, as described by the reference path in fig. (\ref{TSCRGRF}).  The vehicle with initial longitudinal speed  $\dot{x}_p=18$ m/s  is expected to execute a lane change maneuver for avoiding the obstacle and another lane change maneuver to return the vehicle to the reference path again. The set of test cases are designed to pose additional challenge  by adding the yaw rate requirement of a curved road. 

\subsubsection{Test case: Obstacle at Prediction Horizon}
\noindent In this example, there is a detection of the obstacle at the prediction horizon of the controller, i.e., $t=2$ s, on a curved road. The results illustrated by fig. (\ref{TSCRLE}-\ref{TSCRSenv}) reveal that both DMPC and RMPC successfully are able to evade the obstacle. The performance of the DMPC is poor, especially, in the post avoidance phase as shown in fig. (\ref{TSCRLE}) , where the overshoot around the path distance $s_d=80$ m is more than $40$ cm. The controllers  shown in fig. (\ref{TSCRSenv}) are successful in keeping the vehicle inside the imposed stability constraints. The trajectories are closer to the yaw rate bound constraints due to curved road requirement. Overall, due to early detection, both the controllers are able to plan a safe trajectory around the obstacle. 

\begin{figure}
   \centering
        \begin{subfigure}[t]{0.45\textwidth}
        \includegraphics[height=1.8in,width=\textwidth]{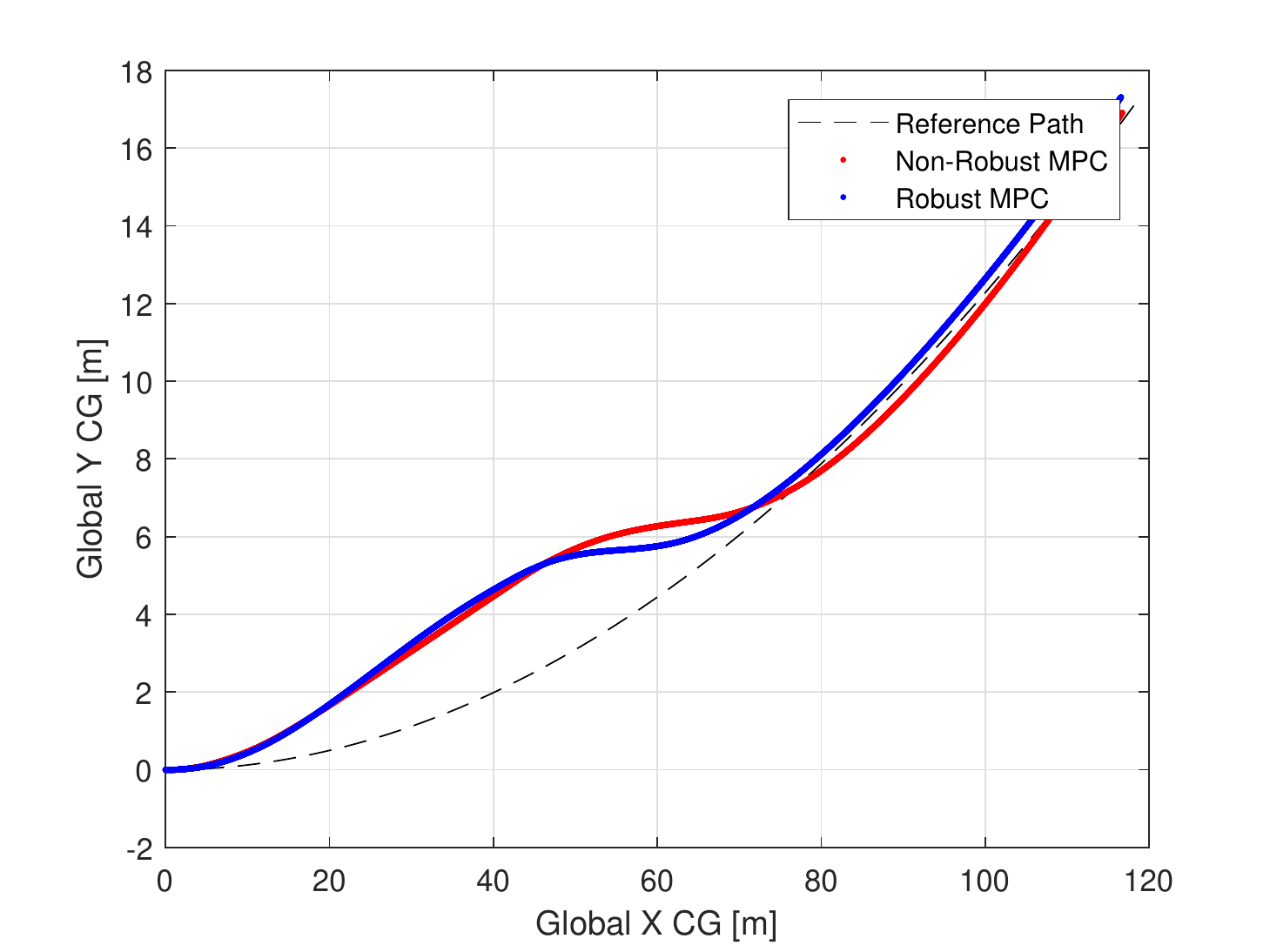}
        \caption{Global reference frame}
        \label{TSCRGRF}
        \end{subfigure}
    \end{figure}
    \begin{figure}\ContinuedFloat
    \vspace*{-0.5cm}
    \centering
      \begin{subfigure}[t]{0.45\textwidth}
        \includegraphics[height=1.8in,width=\textwidth]{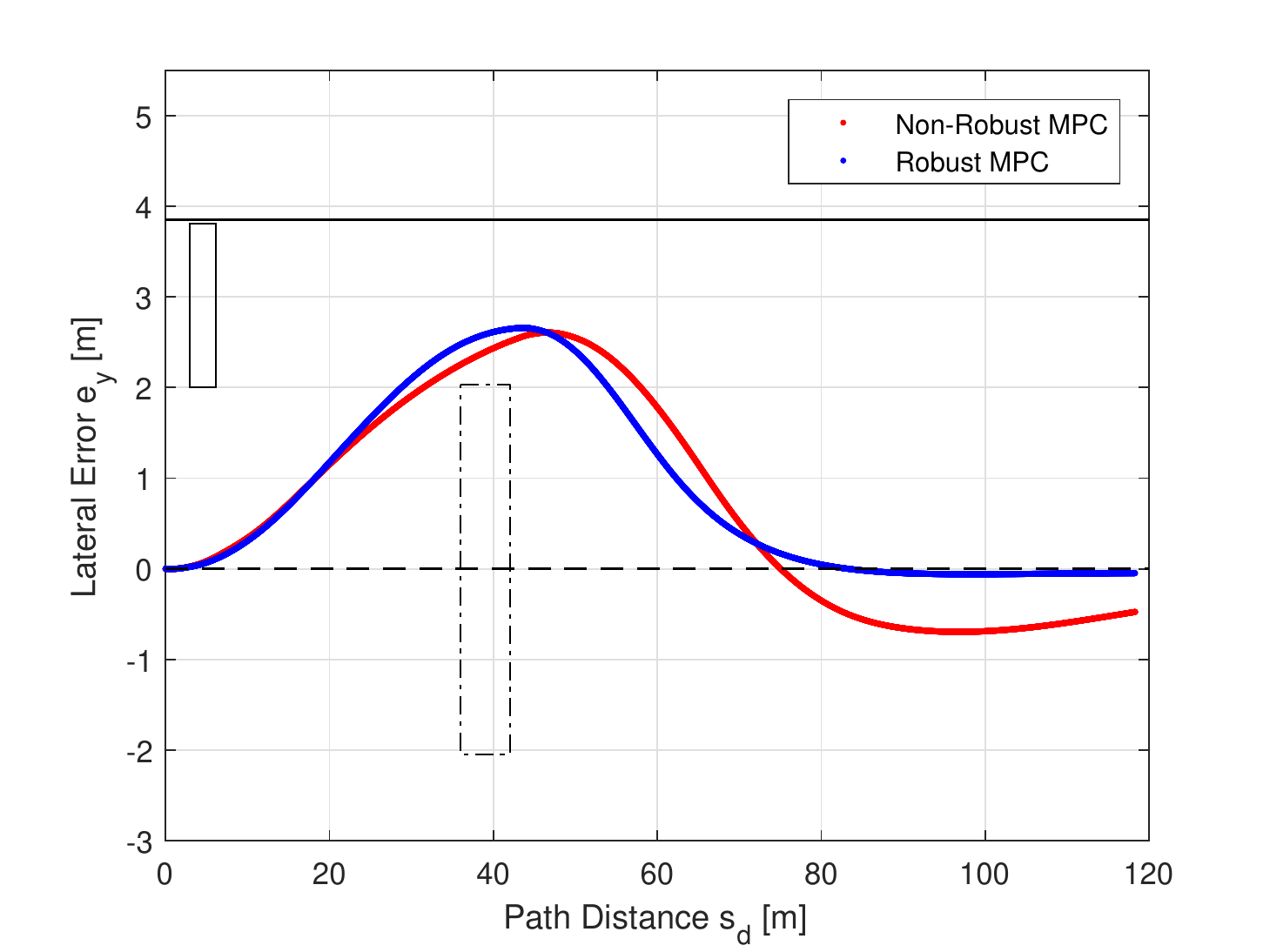}
        \caption{Lateral error}
        \label{TSCRLE}
    \end{subfigure}
     \begin{subfigure}[t]{0.45\textwidth}
        \includegraphics[height=1.8in,width=\textwidth]{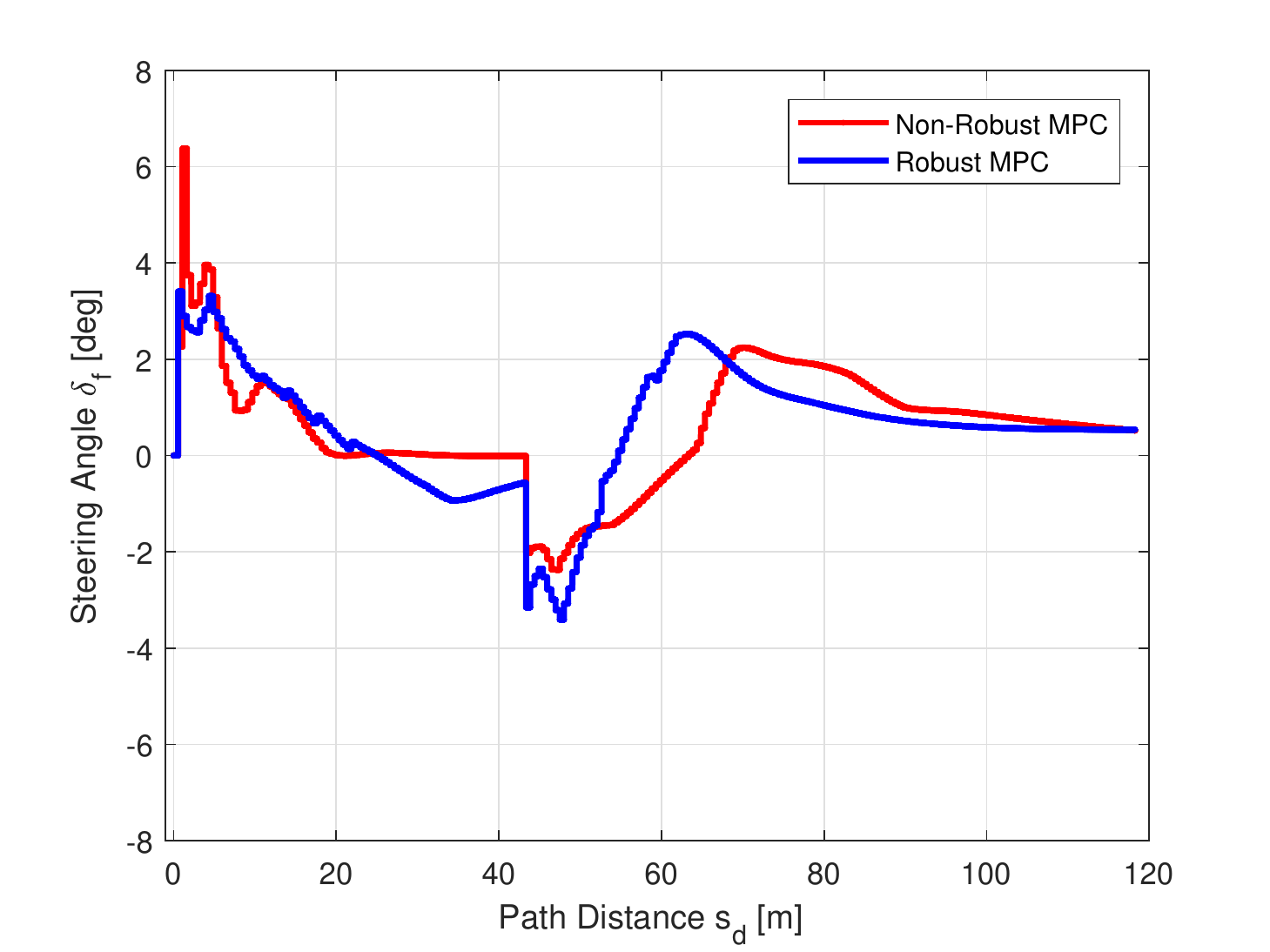}
        \caption{Steering angle input}
        \label{TSCRSteang}
    \end{subfigure}
 \begin{subfigure}[t]{0.45\textwidth}
        \includegraphics[height=1.8in,width=\textwidth]{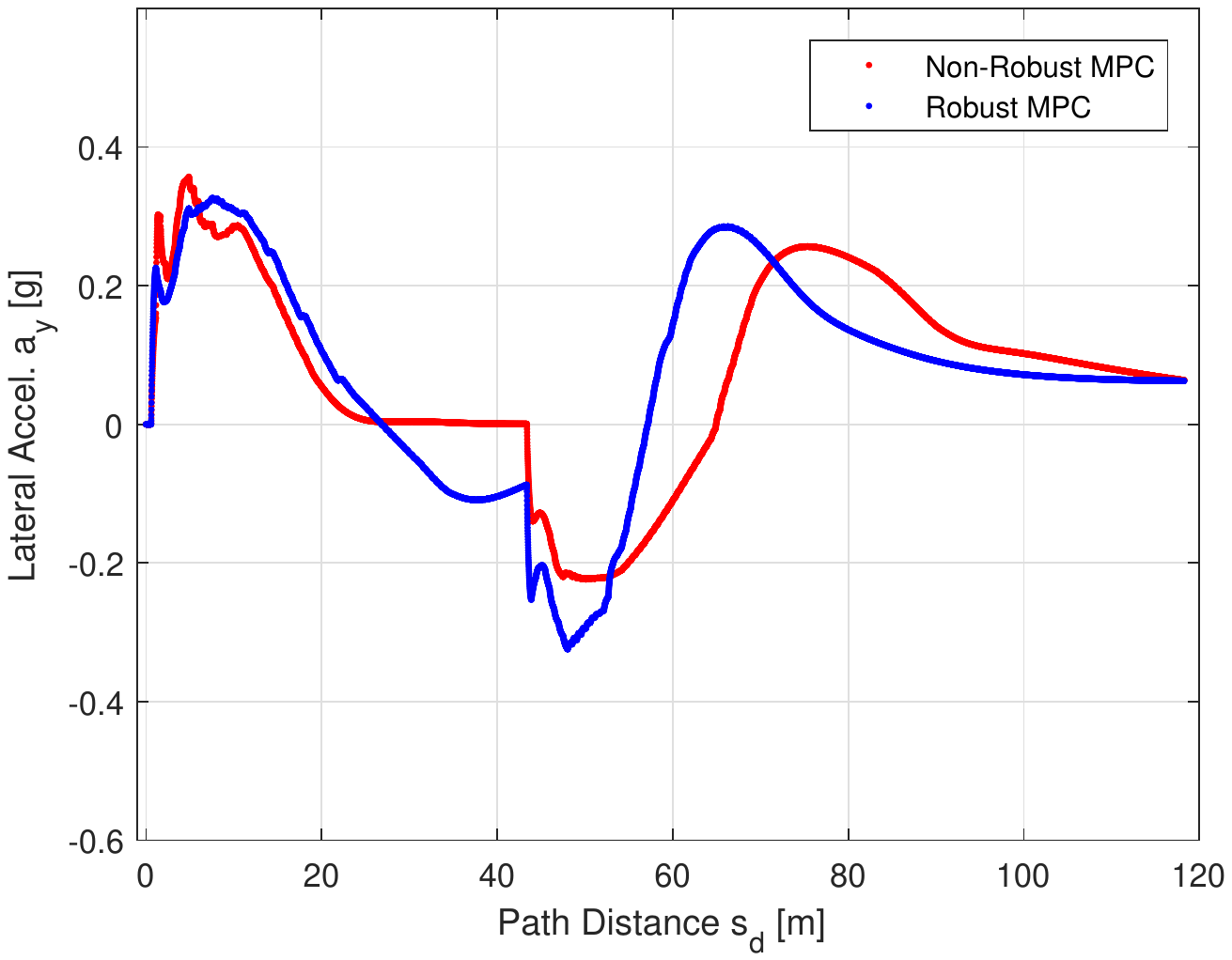}
        \caption{Lateral acceleration}
        \label{TSCRLatacc}
    \end{subfigure}
   \begin{subfigure}[t]{0.45\textwidth}
        \includegraphics[height=1.8in,width=\textwidth]{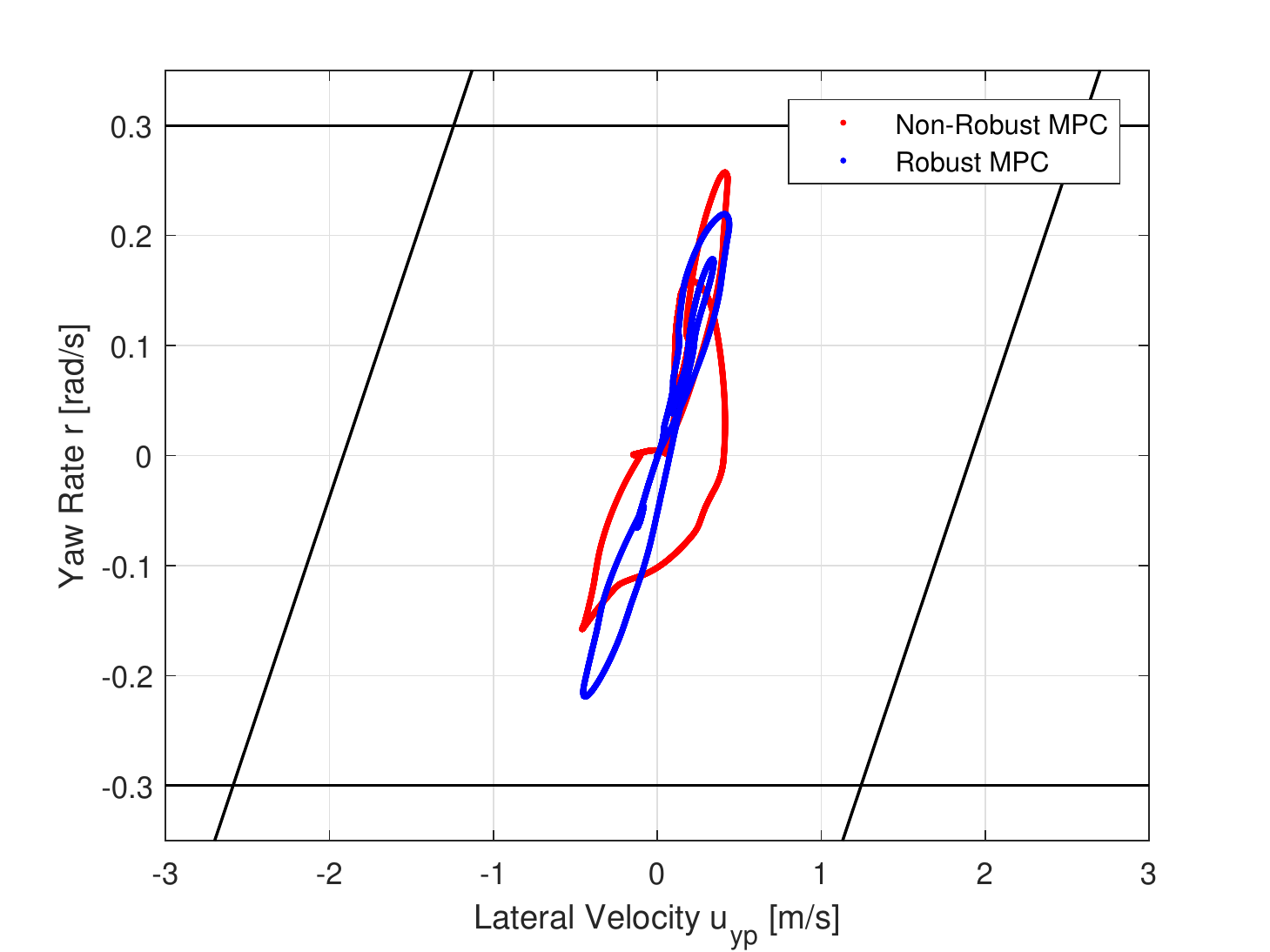}
        \caption{Closed loop trajectories in yaw rate v/s lateral velocity (CP) phase plane with stability constraint}
        \label{TSCRSenv}
    \end{subfigure}
    \caption{Test case: Curved road, $\dot{x}_p = 18 $ m/s, obstacle at prediction horizon. Both controllers are able to avoid the obstacle and return to the prescribed path because of early obstacle detection}\label{TSCR}
\end{figure}

\subsubsection{Test case: Pop-up obstacle }
\begin{figure}
\vspace*{-0.5cm}
\centering
     \begin{subfigure}[t]{0.45\textwidth}
        \includegraphics[height=1.8in,width=\linewidth]{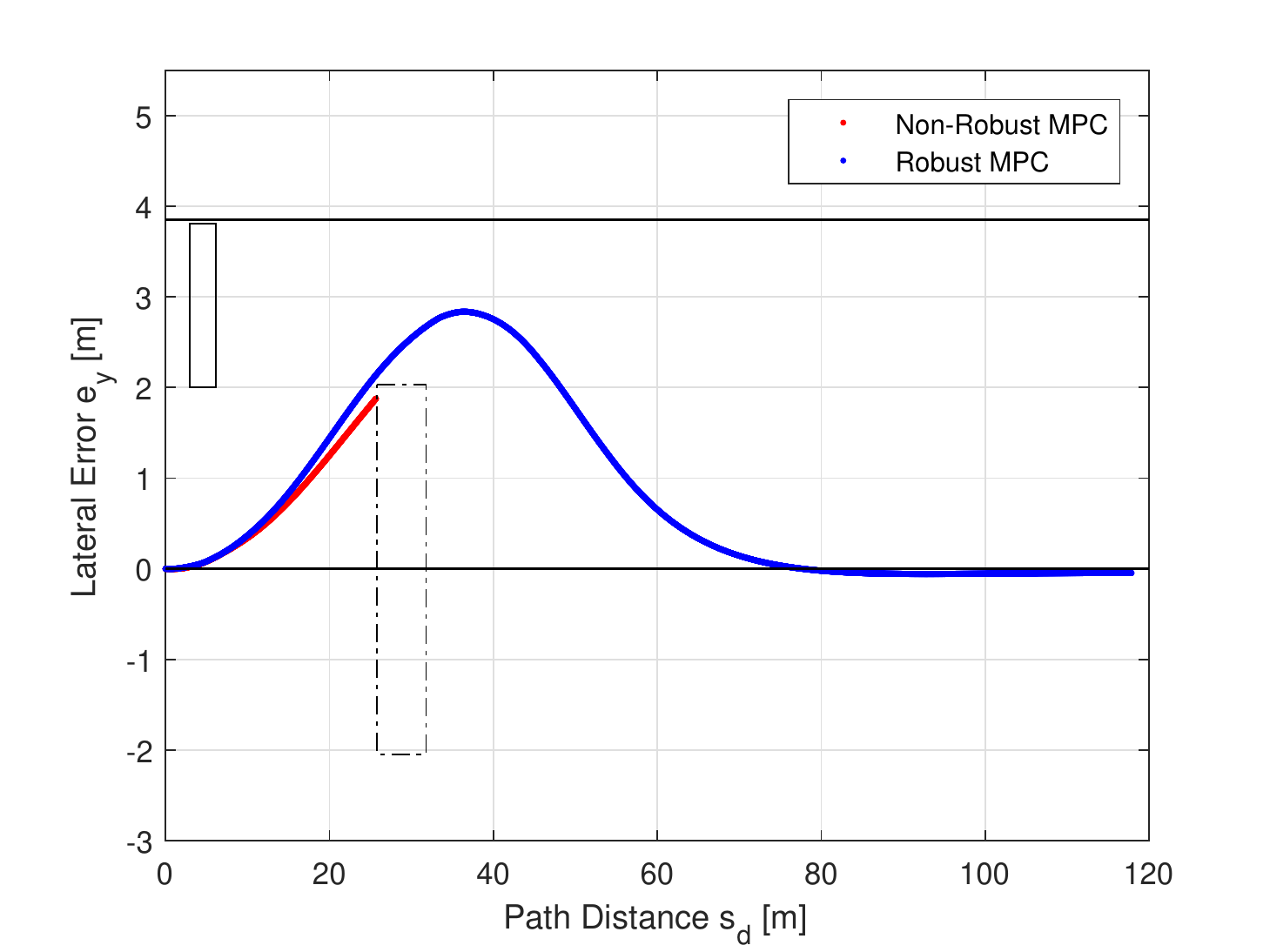}
        \caption{Lateral error}
        \label{TSCRPopUpLE}
    \end{subfigure}
    \begin{subfigure}[t]{0.45\textwidth}
        \includegraphics[height=1.8in,width=\textwidth]{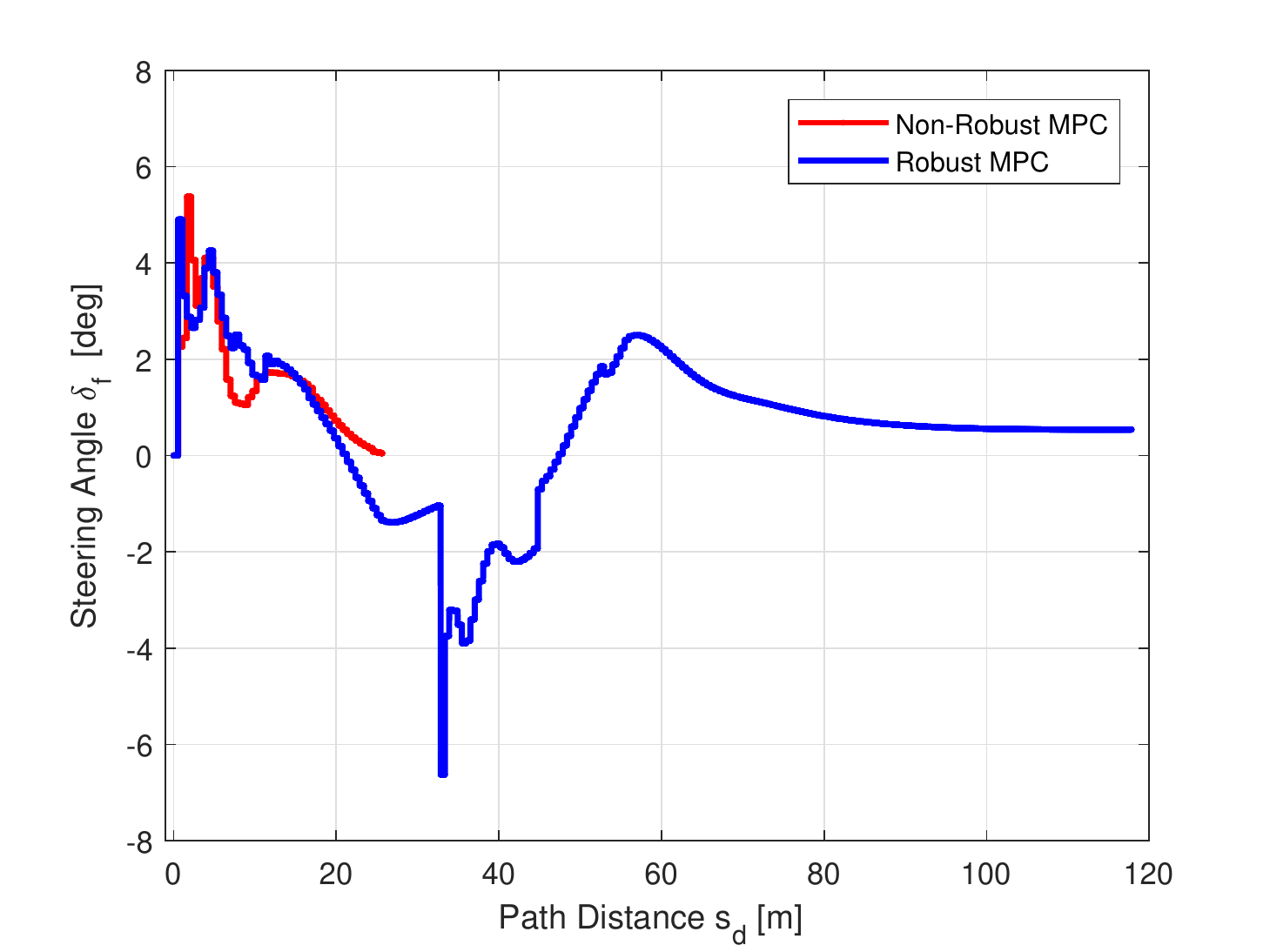}
        \caption{Steering angle input}
        \label{TSCRPopUpSteang}
    \end{subfigure}
   \begin{subfigure}[t]{0.45\textwidth}
        \includegraphics[height=1.8in,width=\textwidth]{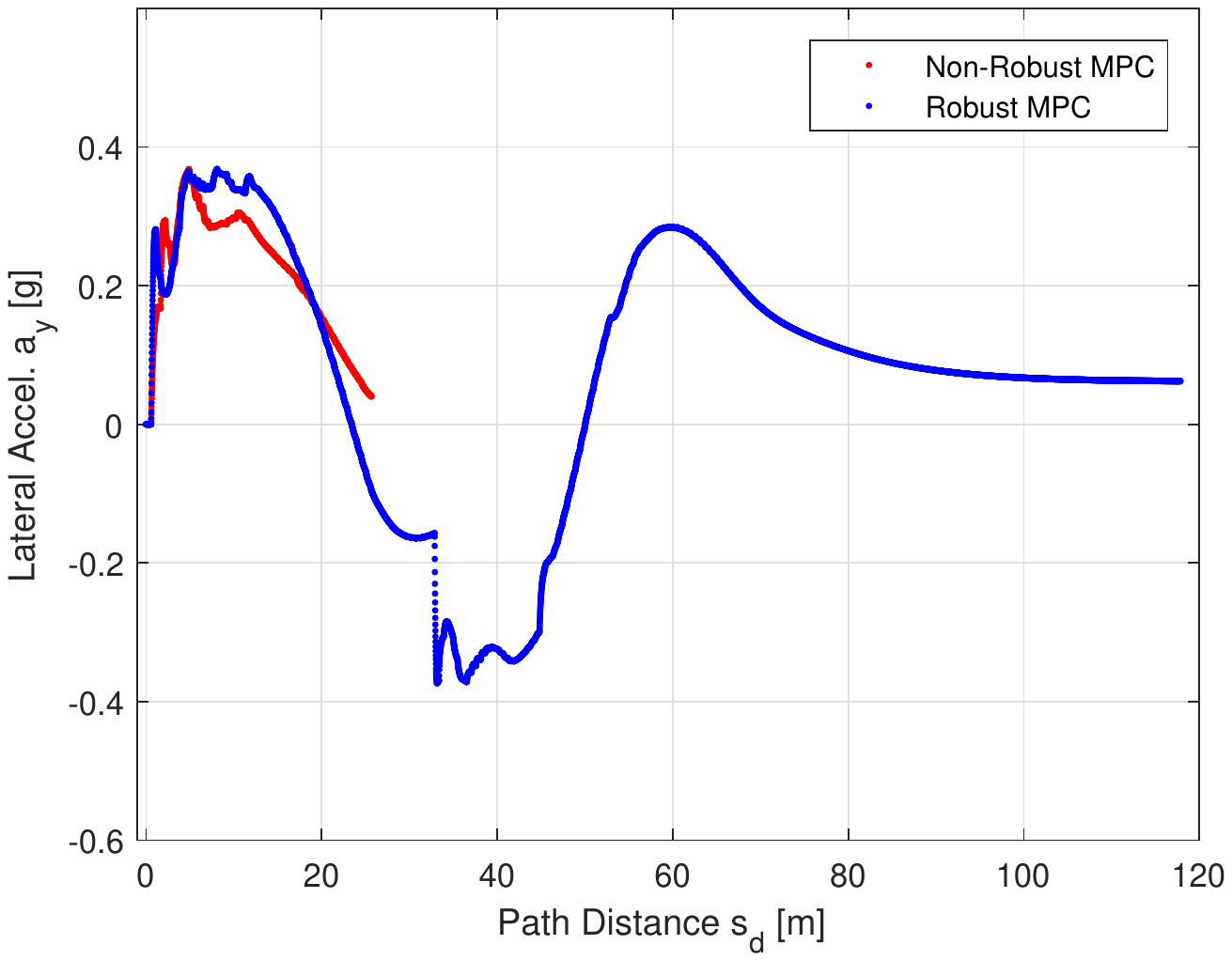}
        \caption{Lateral acceleration}
        \label{TSCRPopUpLatAcc}
    \end{subfigure}
    \begin{subfigure}[t]{0.45\textwidth}
        \includegraphics[height=1.8in,width=\textwidth]{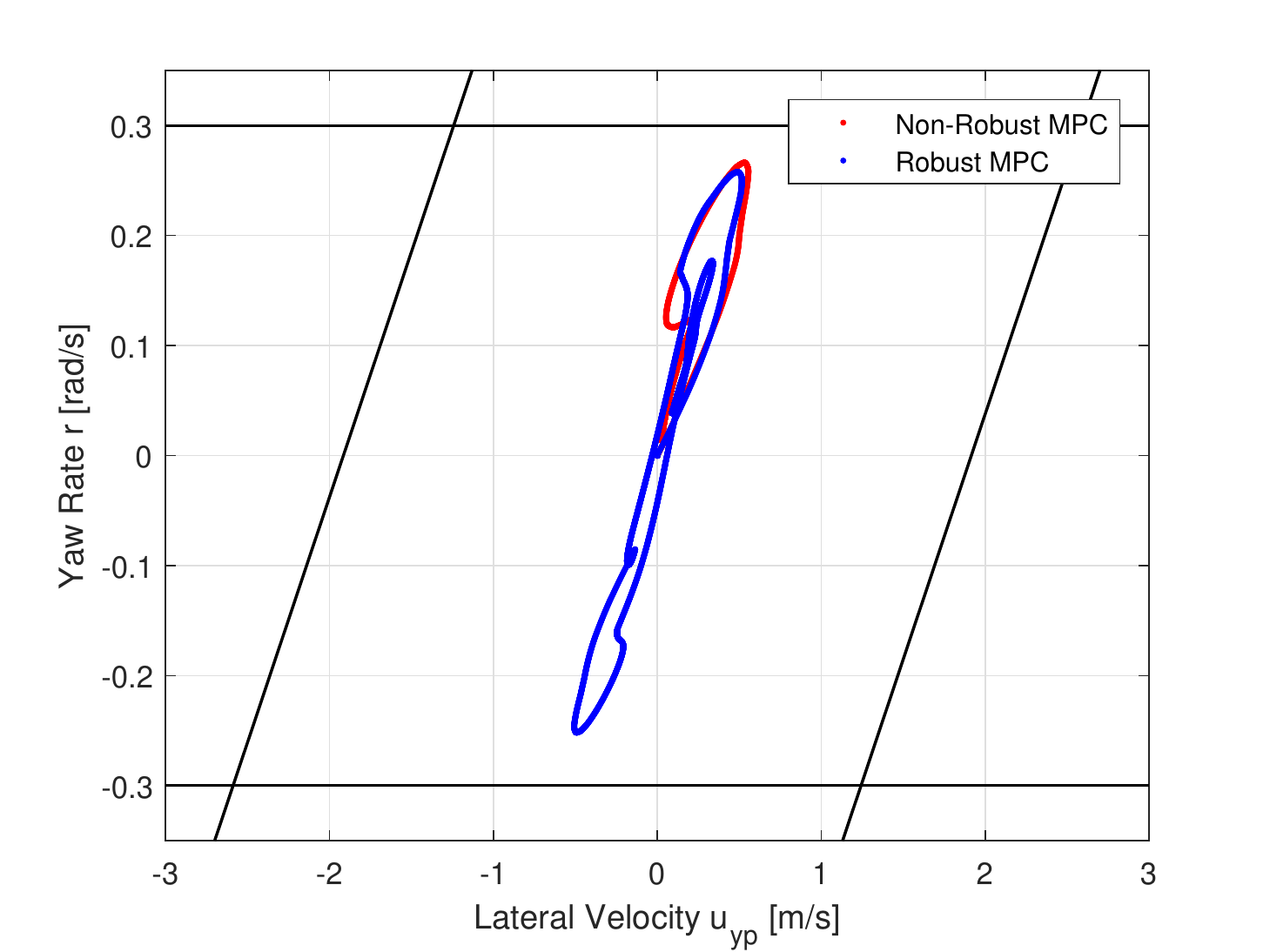}
        \caption{Closed loop trajectories in yaw rate v/s lateral velocity (CP) phase plane with stability constraint}
        \label{TSCRPopUpSenv}
    \end{subfigure}
    \caption{Test case: Curved road, $\dot{x}_p = 18$ m/s, pop-up obstacle at $t=1.4$ s. RMPC controller able to avoid the obstacle and return to the prescribed path, whereas, DMPC failed to find a collision free path.}\label{TSCRPopUp}
\end{figure}

\noindent This test case is an extension of the previous test case which represents either a case of delayed detection or a pop up obstacle in the reference path of the vehicle. Fig. (\ref{TSCRPopUp}) shows the resulting lateral errors, steering angle input, lateral acceleration and phase plane trajectory. Similar to the reasons described in pop-up obstacle on a straight road, the DMPC fails to avoid the obstacle and resulted in a collision. On the other hand, the RMPC controller managed to avoid the obstacle. The difference is the consideration of the model uncertainty in RMPC, resulting in necessary adjustment to the planned trajectory. However, this test case demonstrates the limitations of an obstacle avoidance and path tracking controller system relying only on steering actuation because the trajectory executed by RMPC could only manage to avoid the obstacle by a small gap, as shown in fig. (\ref{TSCRPopUpLE}). This is understood as the artifact of not being able to sustain the high lateral acceleration for the initial path distance ($s_d=[5-20]$ m), as shown in fig. (\ref{TSCRPopUpLatAcc}), due to predicted impending violation of the yaw rate bound, described in fig. (\ref{TSCRPopUpSenv}).

\subsubsection{Test case: Pop-up Obstacle with disturbance}
\begin{figure}

   \centering
    \begin{subfigure}[t]{0.45\textwidth}
        \includegraphics[height=1.8in,width=\linewidth]{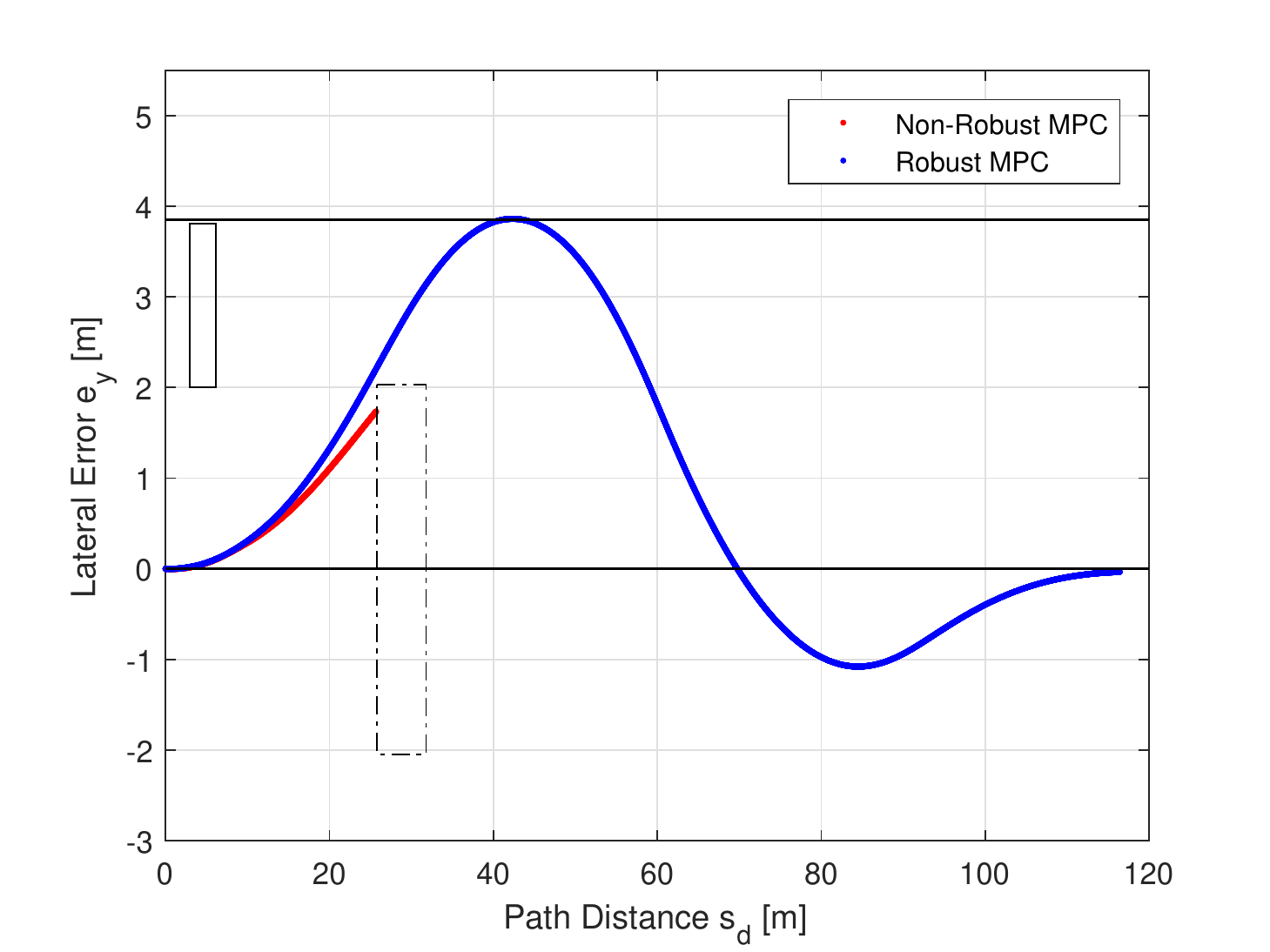}
        \caption{Lateral error}
        \label{TSCRUncerFricLE}
    \end{subfigure}
    \begin{subfigure}[t]{0.45\textwidth}
        \includegraphics[height=1.8in,width=\textwidth]{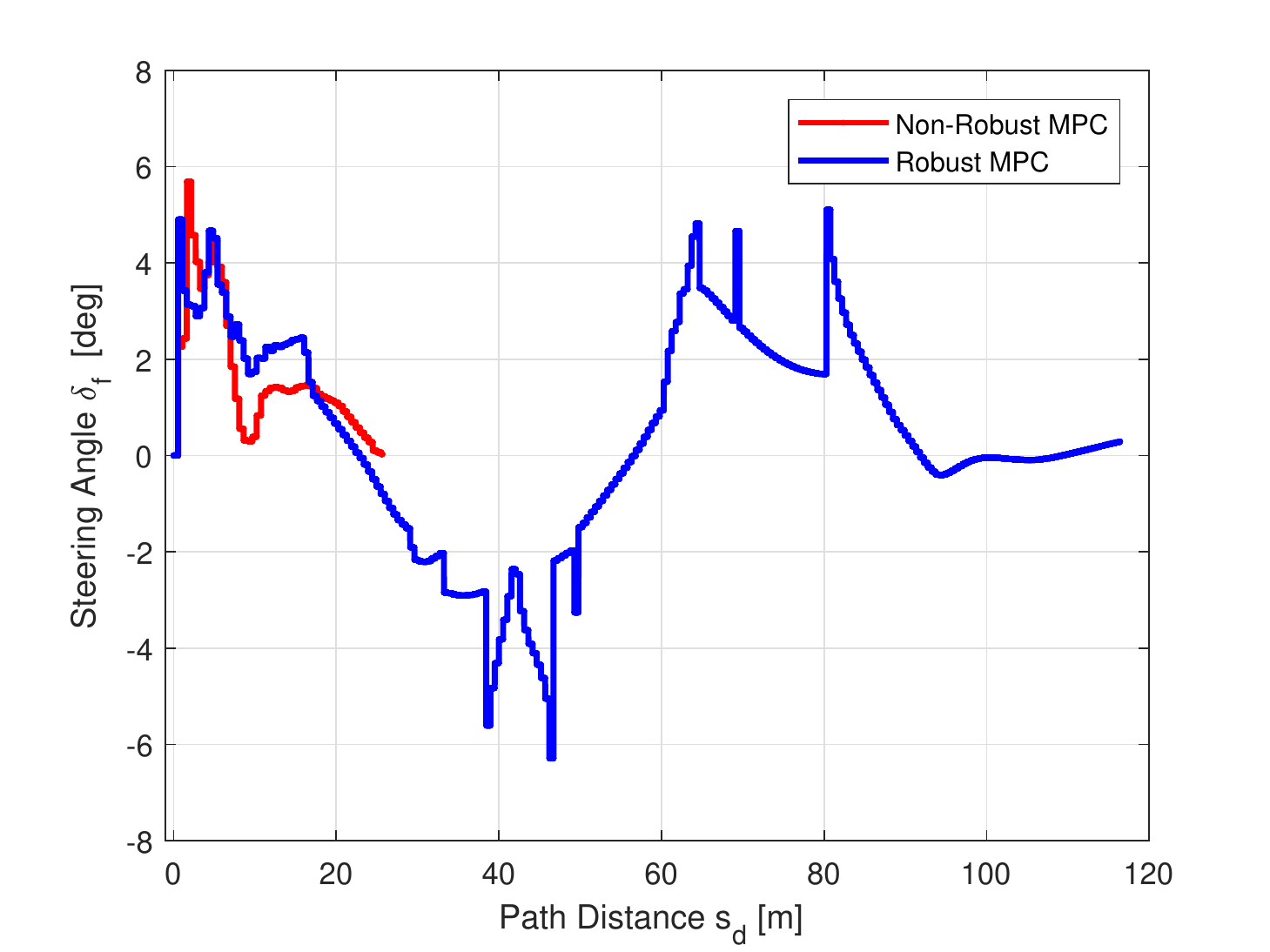}
        \caption{Steering angle input}
        \label{TSCRUncerFricSteang}
    \end{subfigure}
   \begin{subfigure}[t]{0.45\textwidth}
        \includegraphics[height=1.8in,width=\textwidth]{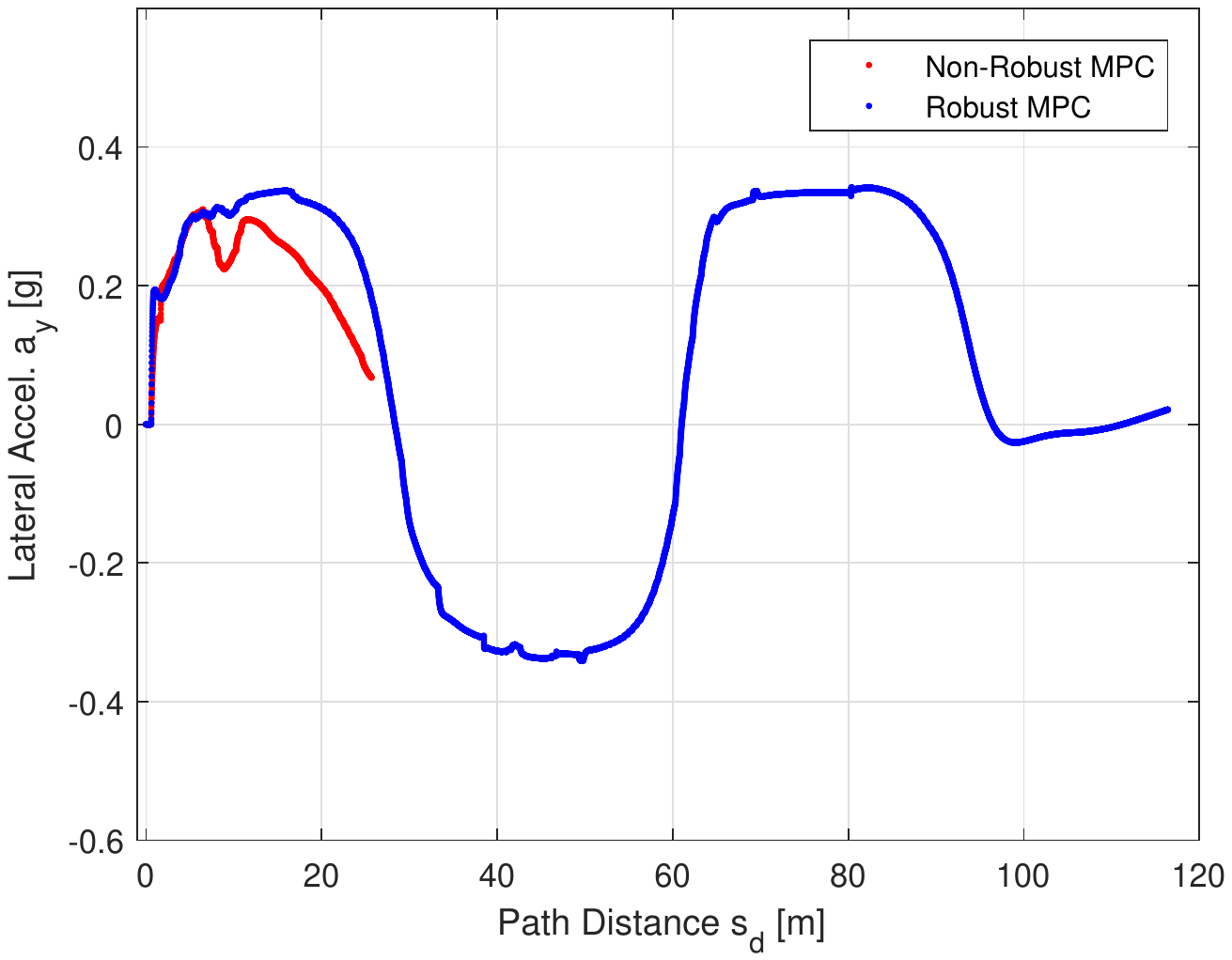}
        \caption{Lateral acceleration}
        \label{TSCRUncerFricLatacc}
    \end{subfigure}
   \begin{subfigure}[t]{0.45\textwidth}
        \includegraphics[height=1.8in,width=\textwidth]{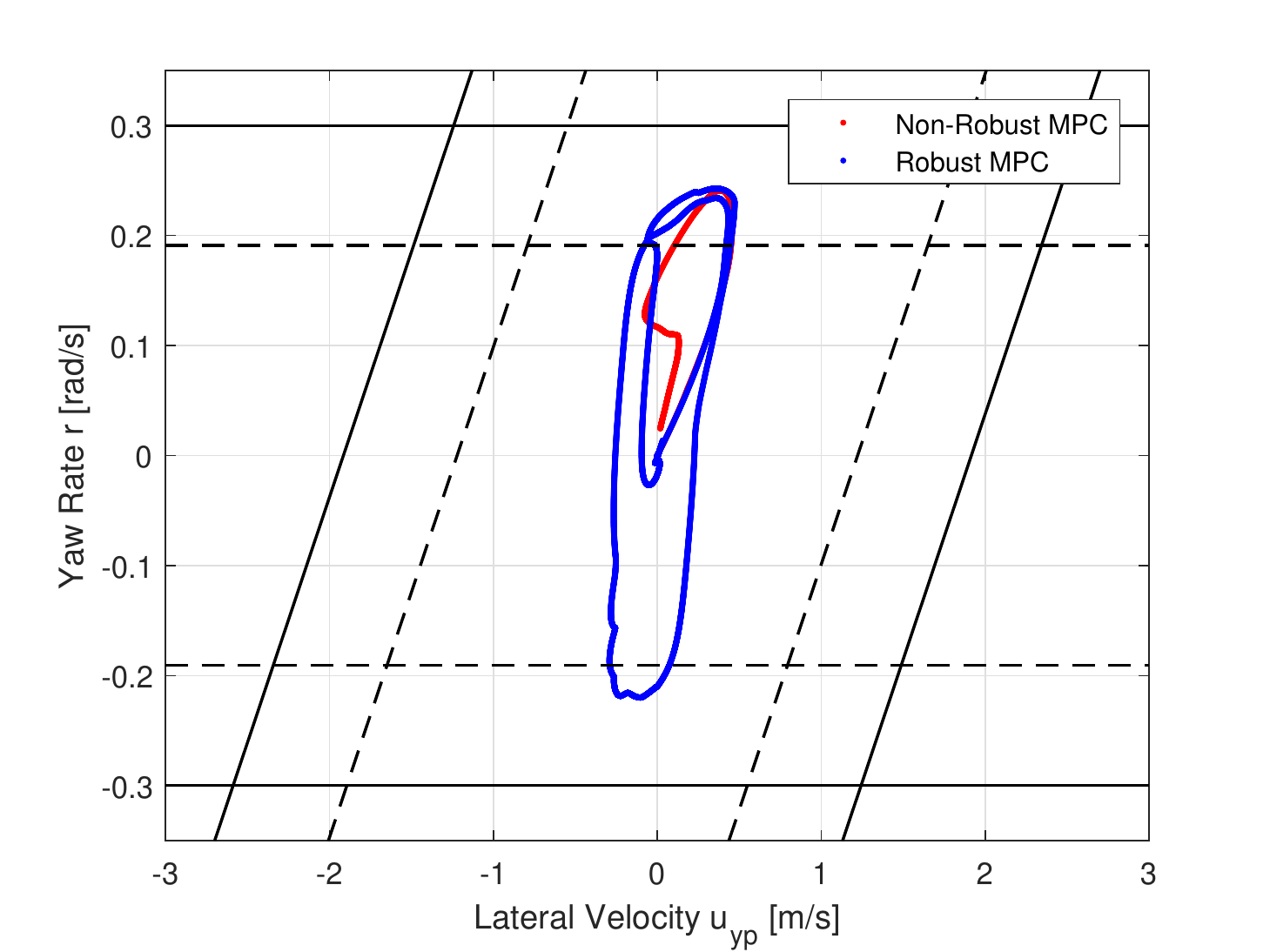}
        \caption{Closed loop trajectories in yaw rate v/s lateral velocity (CP) phase plane, solid lines represents controller stability constraint, and dotted lines represents actual constraints}
        \label{TSCRUncerFricSenv}
    \end{subfigure}
    \caption{Test case: Straight road, $\dot{x_p} = 18 $ m/s, pop-up obstacle at $t=1.4$ s, $\mu_{act} = 0.35$ and $\mu_{c} = 0.55$. RMPC able to avoid the obstacle but failed to satisfy the road boundary and overshoot requirement. DMPC failed to find a collision free path.}\label{TSCRUncerFric}
\end{figure}

\noindent Fig. (\ref{TSCRUncerFric}) shows the results of the pop-up obstacle test case on a curved road. The controllers are set up for the nominal $\mu_c \approx 0.55$ and the tire-road friction coefficient for the CARSIM vehicle solvers is set around $\mu_{act}=0.35$. The vehicle for RMPC is able to avoid the obstacle and track the center line of the road at the speed of $18 $ m/s, however, barely able to keep the vehicle within the safe road boundary limits, as shown in fig. (\ref{TSCRUncerFricLE}). In fig. (\ref{TSCRUncerFricLE}), controller RMPC shows considerable overshoot $s_d=85$ m, after avoiding the obstacle. This is because the RMPC maintained high steering command for longer than in previous test case to compensate for the reduced friction/lateral acceleration, as shown in fig. (\ref{TSCRPopUpLatAcc}) and fig. (\ref{TSCRUncerFricSteang}) for $s_d=[5-20]$ m. 

In a similar case with lower vehicle longitudinal speed, this would have resulted in less path distance covered for the same time duration required and could have resulted in steering angle reduction at path distance earlier than for the current velocity case leading to maneuver not close to the road boundary limit. Therefore, reduction in longitudinal velocity will be explored in future work by including the brake actuation and the result will be compared to the result for the integrated braking and steering actuated control system. 

\section{Conclusion}

\noindent This paper presents a fast-online robust MPC framework based lateral control system for obstacle avoidance and path tracking of autonomous vehicle. A discrete linear time varying using successive linearization of a non-linear tire model is  developed and used in the control design. A novel method of incorporating the error reachable sets information to a discrete convex search space is developed. A tube-based MPC is used with tightened search space to ensure obstacle avoidance and path tracking in the presence of unknown disturbances. Multiple simulations with different test cases have been run to evaluate the controller performance and its comparison to a deterministic linear MPC controller. The results highlighted the importance of robust design for planning and execution of emergency maneuvers and for path tracking in slippery conditions for autonomous vehicles.


%





\ifCLASSOPTIONcaptionsoff
  \newpage
\fi

\bibliographystyle{IEEEtran}
\bibliography{Lateral_Control}

\begin{IEEEbiographynophoto}{Vivek Bithar}
\newline Vivek Bithar received the B.S. degree in production and industrial engineering from the Delhi College of Engineering, Delhi, India, in 2009, the M.S. degree in mechanical engineering from University of Michigan, Ann Arbor, MI, USA, in 2010, and the Ph.D. degree in mechanical engineering from the Ohio State University in 2020, with a focus on robust optimal predictive control systems and its application to obstacle avoidance for automated vehicles.
\end{IEEEbiographynophoto}

\begin{IEEEbiographynophoto}{Punit Tulpule}

\newline Punit Tulpule received his Bachelors of Science degree in Physics and Master of Science degree in Atmospheric Sciences from University of Pune, India in 2006 and 2008 respectively. He received PhD degree in Mechanical Engineering from Iowa State University in 2014. He is currently a Research Assistant Professor of Mechanical and Aerospace Engineering Department at the Ohio State University (OSU). He has held different researcher positions at Simulation Innovation and Modeling Center and Center for Automotive Research at OSU. His research interests are in validation and verification of complex engineering systems, modeling and simulation of large scale systems, and optimal control.  
\end{IEEEbiographynophoto}

\begin{IEEEbiographynophoto}{Shawn Midlam-Mohler}
\newline Shawn Midlam-Mohler received his bachelors, masters and PhD degrees from The Ohio State University (OSU) in 1999, 2001 and 2005 respectively. He has held different research faculty and staff positions at OSU's Center for Automovite Research until 2014. Currently he is the Director of Simulation Innovation and Modeling Center and professor of practice in the Mechanical and Aerospace Engineering Department.  His research interests are in the area of design, system modeling, applied controls, and systems engineering.  He has over one hundred patents and peer-reviewed publications and his main research focus is in the area of advanced automotive systems with a focus on energy and emissions.  
\end{IEEEbiographynophoto}

\end{document}